\documentclass[a4paper,11pt]{article}
\usepackage{jcappub} 
\usepackage{color}
\usepackage[dvipsnames]{xcolor}
\usepackage{lineno}
\usepackage{booktabs}
\usepackage{bbding}
\usepackage{tabularx}
\usepackage{subcaption}
\usepackage{multirow}
\usepackage{array}
\usepackage{float}

\arxivnumber{2507.13890}
\title{Investigating $f(R)$-Inflation: background evolution and constraints}
\author[1, 2, 3]{E. Fazzari\thanks{},}
\author[1,2,3,4]{C. De Leo\thanks{},}

\author[1,5]{G. Montani,}
\author[2,4]{M. Martinelli,}
\author[1, 2]{A. Melchiorri,}
\author[6, 7]{G. Cañas-Herrera.}

\affiliation[1]{\itshape Physics Department, Sapienza University of Rome, P.le A. Moro 5, 00185 Roma, Italy}
\affiliation[2]{\itshape Physics Department, INFN - Sezione di Roma, P.le A. Moro 5, I-00185, Roma, Italy}
\affiliation[3]{\itshape Physics Department, Tor Vergata University of Rome, Via della Ricerca Scientifica 1, Roma, 00133, Italy}
\affiliation[4]{\itshape INAF - Osservatorio Astronomico di Roma, via Frascati 33, 00078 Monte Porzio Catone, Italy}
\affiliation[5]{\itshape Nuclear Department, ENEA - C.R. Frascati, Via E. Fermi 45, 00044 Frascati, Italy}
\affiliation[6]{ESTEC - European Space Agency, Keplerlaan 1, 2201 AZ Noordwijk, The Netherlands}
\affiliation[7]{Leiden Observatory, Leiden University, Niels Bohrweg 2, 2333 CA, Leiden, the Netherlands}
\footnotetext{These authors contributed equally to this work}
\date{\today}

\emailAdd{elisa.fazzari@uniroma1.it}
\emailAdd{chiara.deleo@uniroma1.it}
\emailAdd{giovanni.montani@enea.it}
\emailAdd{matteo.martinelli@inaf.it}
\emailAdd{alessandro.melchiorri@uniroma1.it}
\emailAdd{canasherrera@strw.leidenuniv.nl}

\abstract{
In this work, we investigate the possibility of generating an inflationary mechanism within the framework of a metric-$f(R)$ modified gravity theory,  formulated in the Jordan frame. We explore whether the scalar field, non-minimally coupled to gravity and emerging in the Jordan frame, can play the role of the primordial inflaton. Particular attention is devoted to constructing a dynamical scenario in the Jordan frame that exhibits a slow-rolling phase for the scalar field and admits a quasi-de Sitter solution for cosmic evolution. To ensure consistency with the standard cosmological model, we impose a matching condition with the $\Lambda$CDM model at the end of the inflationary phase. Furthermore, to address the problem of the absence of matter after inflation, we consider a radiation-type particle creation process that maintains an approximately constant energy density. We test our theoretical model against background observational data, specifically Pantheon$^+$ calibrated with SH0ES and DESI calibrated with BBN. We asses the model's viability by combining theoretical consistency tests with its predictions for primordial power spectrum observables, and we discuss the implications for alleviating the Hubble constant tension.
}

\begin{document}
\maketitle
\flushbottom

\section{\label{sec:Introduction} Introduction}
First proposed in the early 1980s \cite{Starobinsky, Sato:1981qmu, Guth, LINDE} as a solution to fundamental paradoxes of the Big Bang framework, the inflationary scenario has since become a pillar of modern cosmology \cite{KolbTurner, weinberg, Primordial}. Its success lies not only in resolving issues such as the horizon and monopole problems, but also in making testable predictions that have been strongly supported by observational data. Measurements of the cosmic microwave background reveal that today’s spatial geometry is remarkably flat \cite{boomerang, efstathiou_planck} (see also \cite{divalentino_planck}), and they confirm a predominantly scale-invariant spectrum of primordial fluctuations \cite{Planck18}. Although we do not yet have a definitive proof of inflation \cite{riotto2017lectures}, the evidence strongly suggests that the early Universe underwent an inflationary phase, and a broad class of viable inflationary models continues to be explored.
\newline
Among the most compelling theoretical motivations for inflation are those rooted in extensions of the Standard Model of particle physics, most notably Grand Unified Theories (GUTs), which naturally predict an early phase dominated by vacuum energy density \cite{KolbTurner, peacock_cosmology}. By combining together fundamental particle interactions and cosmic evolution, GUTs offer a concrete foundation for the inflationary paradigm. Yet implementations of inflation driven by GUT symmetry breaking often feel \emph{ad hoc}, invoking specific assumptions about the symmetry-breaking transition and require fine-tuned parameters to sustain a sufficiently long de Sitter–like expansion \cite{review_particlephysics}. The need for precise parameter choices partly reflects the fine-tuning of the Universe’s initial conditions, which underlies the fundamental paradoxes that inflation’s de Sitter phase and subsequent reheating were meant to resolve \cite{weinberg}.

Because particle-physics–based models often feel ad hoc, researchers have turned to modified gravity theories beyond the Standard Model \cite{Nojiri-odintsov, Capozzielo-Delaurentis, nojiri_2017}. The first, and probably most successful, approach to interpret inflation via a modified gravity paradigm was proposed by Starobinsky \cite{Starobinsky, Vilenkin:1985md}. This formulation relies on a quadratic correction in the Ricci scalar to the Einstein-Hilbert action for gravity.  In the very early Universe, when spacetime curvature was extremely high, this correction naturally drives a period of exponential expansion. As the Universe expands and curvature drops, the 
quadratic term quickly becomes negligible, smoothly ending inflation and transitioning into the standard hot Big Bang phase. For possible extensions to the Starobinsky model see \cite{Nearly_Starobinsky, Ivanov:2021chn, Appleby:2009uf, Comments_Starobinsky, Kuralkar:2025zxr, Odintsov:2026fyc}.
\newline
The present analysis adopts Starobinsky’s inflationary framework but with a different approach: rather than assuming a specific $f(R)$-Lagrangian a priori (see \cite{Sotiriou_rev}), we work in the Jordan frame and treat the scalar-field potential (from the non-minimal coupling) as a fully dynamical quantity. Only after solving the dynamics, we reconstruct the corresponding $f(R)$ form \emph{a posteriori}. This reconstruction is intended to be valid only within the specific evolutionary regime under consideration. For a similar late-Universe implementation, see \cite{montani_mary_fR, matcre_montanimary, MONTANI_mary}. For a comprehensive review about $f(R)$ gravity inflation and other related works see \cite{Odintsov:2023weg, Oikonomou:2025qub, Odintsov:2020thl, Oikonomou:2020oex, Lopez:2025gfu, Kouniatalis:2025orn, Gomes:2016cwj, Gomes:2018uhv}. 

Our aim is to test the viability of this inflationary scenario and its implications for background data. In particular, we assess whether it can alleviate the \emph{Hubble tension} \cite{whitepaper_cosmoverse,divalentino-review,montani_hubbletension_1,Montani_hubbletension_2,montani_hubbletension_4,montani_hubbletension_6,montani_mary_fR,montani_hubbletension_8,giare_inflation}. To that end, we first derive revised inflationary dynamics in the Jordan frame, based on modified gravity and the hypothesis that a rapidly changing gravitational field produces radiation-like matter \cite{matcre_calvaoLima,matcre_montani2001,fazzari_H0z}, providing a natural reheating mechanism. We then compare the model against
DESI data (with BBN calibration) \cite{DESIDR1,desicollaboration2025desidr2resultsii} and the Pantheon$+$ sample (with SH0ES calibration) \cite{Riess_2022}.

The structure of the paper is as follows. In \autoref{sec:theory} we present our modified gravity Inflation model, starting from introducing the modified $f(R)$-gravity scenario with the theoretical constraints for a viable $f(R)$ theory \autoref{subsec:viability_fR}, and the radiation-type particle creation mechanism \autoref{subsec:particle_cre} as the reheating process. In \autoref{sec: formulation} we show the dynamical formulation of the proposed inflationary theory.  
Then we move on testing the model at a background level in \autoref{sec: model_testing} against current data. We present the results on the simulations side in \autoref{sec:results} and the theoretical implications in \autoref{sec:theo_results}, where we also provide predictions for primordial power spectrum observables. Finally, we draw our main conclusions in \autoref{sec:conclusion}.

\section{Modified gravity inflationary model \label{sec:theory}}

In this section, we present the theoretical foundation of our inflationary model, beginning with the modified gravity scenario in the Jordan frame, followed by the particle creation mechanism, the dynamical equations, and a theoretical analysis of the fiducial parameter values used for model testing.

\subsection{Modified gravity theory: $f(R)$-metric scenario in the Jordan frame \label{subsec: f(R)}}
\noindent $f(R)$ modified gravity models \cite{Sotiriou_rev, Beyond, Nojiri-odintsov} generalize Einstein's General Relativity (GR) by altering the functional form of the Ricci scalar $R$ in the gravitational action. This modification is encapsulated by the function $f(R)$, which incorporates an additional degree of freedom compared to GR. This extra degree of freedom is commonly used to model new phenomena that have emerged in cosmology.\\
The total action for $f(R)$ gravity is, therefore, a generalization of the Einstein-Hilbert action with the usual matter term, as follows:

\begin{equation}
S[g_{\mu\nu}]=\frac{1}{2\chi}\int_{\mathcal{M}}\mathrm{d}^{4}x\sqrt{-g}f(R) + S_{M}[g_{\mu\nu},\psi] \, ,
\label{eq:azione_f(R)mat}
\end{equation}
where, adopting the metric signature $(-,+,+,+)$, $\chi \equiv 8 \pi G$ is the Einstein constant (in natural units), $G$ is the gravitational constant, $g$ is the determinant of the metric tensor $g_{\mu\nu}$, $S_M$ is the matter action, and $\psi$ stands for matter fields. We stress that the action of matter maintains minimal coupling with the metric, as in GR.

Defining the scalar field as $\phi \equiv f'(R) \equiv df/dR$ and the scalar field potential $V(\phi) \equiv \phi R(\phi)-f(R(\phi))$, if $f''(R) \neq 0$, this action in Eq.~\eqref{eq:azione_f(R)mat} can be rewritten in its dynamically equivalent formulation in the so-called Jordan frame \cite{Sotiriou}, taking the form:

\begin{equation}
S_{J}[g_{\mu\nu}]=\frac{1}{2\chi}\int_{\mathcal{M}}\mathrm{d}^{4}x\sqrt{-g}\biggl[\phi R-V(\phi)\biggr]+S_M[g_{\mu\nu},\psi] \, ,
\label{eq:azionetot_Jordan}
\end{equation}
where the potential $V(\phi)$ is fixed by the specific functional expression of $f(R)$.\\
Therefore, we work in the scalar–tensor representation in the so-called Jordan frame. In this formulation, the extra degree of freedom encoded in the function $f(R)$ is recast as a scalar field intrinsic to the gravitational theory, allowing us to explain cosmological effects beyond the reach of GR.
It is worth noting that the scalar field $\phi$ is non-minimally coupled to the metric through the Ricci scalar, and due to the fact that it preserves the coupling with the original metric, it can be quantized and interpreted as a massive scalar mode describing the gravitational field.

We now derive the modified cosmological equations by considering a flat, homogeneous, and isotropic Universe described by the Friedmann–Lemaître–Robertson–Walker (FLRW) line element \cite{weinberg}:

\begin{equation}
ds^{2}=-dt^{2}+a^{2}(t) \left(dx^2+ dy^2+ dz^2\right) \;,
\label{eq:FLRW_lineele}
\end{equation}

\noindent where $a(t)$ represents the scale factor, and $t$ denotes synchronous time.

The source terms involved in cosmological dynamics are described through the stress-energy tensor of a cosmological perfect fluid as:

\begin{equation}
T_{\mu\nu}^{\text{p.f.}}=(\rho+P)u_{\mu}u_{\nu}+P g_{\mu\nu}\,,
\label{eq:def_tensoreEI}
\end{equation}
where $\rho=\rho(t)$ represents the total energy density, $P=P(t)$ is the pressure, and $u_{\mu}$ is the four-velocity of the fluid.

By considering the $tt$-component of both the field equations obtained varying Eq.~\eqref{eq:azionetot_Jordan} with respect to $g_{\mu\nu}$  and the stress-energy tensor in Eq.~\eqref{eq:def_tensoreEI}, the result is the following generalized Friedmann equation:

\begin{equation}
H^{2}=\frac{\chi\rho}{3\phi}-H\frac{\dot{\phi}}{\phi}+\frac{V(\phi)}{6\phi} \; ,
\label{eq:Friedmann_mod}
\end{equation}
where $\dot{} = d/dt$, and $H \equiv \dot{a}/a$ is the Hubble parameter. \\
By varying the action in Eq.~\eqref{eq:azionetot_Jordan} with respect to $\phi$, we obtain:

\begin{equation}
R=\frac{dV}{d\phi} \, ,
\label{eq:vincolo_def}
\end{equation}
and, by explicating the curvature scalar in the FLRW metric in Eq.~\eqref{eq:vincolo_def}, we get:

\begin{equation}
\frac{dV}{d\phi} = 6\dot{H} + 12H^2 \, .
\label{eq:vincolo_FLRW}
\end{equation}
These two equations, i.e. Eqs.~\eqref{eq:Friedmann_mod}, \eqref{eq:vincolo_FLRW} are independent, and it is useful to stress that only the scale factor $a(t)$ and the scalar field $\phi(t)$ are the two fundamental degrees of freedom for cosmological implementation, whereas the potential $V(\phi)$ represents the fingerprint of the particular $f(R)$-Lagrangian adopted.

\subsection{Viability tests for the $f(R)$ theory} \label{subsec:viability_fR}
Given a $f(R)$ theory, it is crucial to asses its physical viability. Considering the definition of the $f(R)$ function from Eq.~\eqref{eq:azionetot_Jordan}, that is 
\begin{equation}
    f(R)=R\phi(R)-V(R) \,,
    \label{eq:fR_def}
\end{equation}
it must satisfy the following conditions:
\begin{equation}
    f_R\equiv\frac{df}{dR}>0 \quad \text{and} \quad f_{RR} \equiv \frac{d^2f}{dR^2}>0 \,.
    \label{eq:fR_check_der}
\end{equation}
The first condition ensures that gravity remains attractive and that the graviton is not a ghost \cite{Sotiriou_rev, DeFelice:2010aj}, while the second prevents the Dolgov-Kawasaki instability \cite{Dolgov:2003px, Appleby:2009uf}.
Additionally, to avoid a tachyonic mode for the scalar field \cite{Sotiriou_rev}, one has to require that 
\begin{equation}
     m^2 = \frac{f_R-Rf_{RR}}{3f_{RR}} > 0 \,,
     \label{eq:m2_def}
\end{equation}
where $m^2$ is the squared mass of the non minimally coupled scalar field $\phi$ computed under the assumption of a constant curvature background, i.e. de Sitter or quasi de Sitter. \\
The squared mass can also be written explicitly in terms of the scalar field and the potential in the Jordan frame \cite{Olmo:2005zr}
\begin{equation}
     m^2 = \frac{1}{3} \left(\phi \frac{d^2V}{d\phi^2}-  \frac{dV}{d\phi}\right) \,.
     \label{eq:m2_phi}
\end{equation}

It is worth noting that we construct the $f(R)$ function \emph{a posteriori} since the potential term of the non-minimally coupled scalar field is treated as a dynamical variable. The resulting modified gravity model is therefore valid solely for describing an inflationary scenario and does not need to be compared with solar system tests, which are relevant to the late Universe as its contribution vanishes after the end of the inflationary epoch.

\subsection{Particle creation mechanism} \label{subsec:particle_cre}
A rapidly varying gravitational field, like other dynamic fields, can produce particles as a result of quantum effects induced on microscopic matter fields \cite{Bondi_1948, Hoyle_1948, Lima:1999rt, Singh_2011}. 
In this work, we explore a phenomenological representation within a thermodynamic framework (for similar works see \cite{matcre_calvaoLima, matcre_montani2001, Primordial, Ramos_2014, deHaro:2015hdp, matcre_nunes2015, matcre_nunes2016, elizalde_odintsov, matcre_montanimary, Schiavone:2026agq}).

To allow for particle creation, we consider an open thermodynamical system and deriving the modified continuity equation starting from the first principle of thermodynamics:

\begin{equation}
dU = - P dV + \delta Q + \mu dN,
\label{eq:1_thermo}
\end{equation}
where $U$ is the internal energy, $P$ the pressure, $\mu$ the chemical potential, $N$ the particle number, and $V$ the volume of the considered system. Now, if we consider the second thermodynamics principle $\delta Q= T dS$, rewriting $U = \rho V$, $S = \sigma N$, together with the Gibbs free-energy definition $\mu = \left(\rho + P\right) V / N - T \sigma$,  Eq.~\eqref{eq:1_thermo} can be restated as follows:
\begin{equation}
d\rho = - (\rho + P) \left( 1 - \frac{d \ln N}{d \ln V} \right) \frac{dV}{V} + \frac{T}{N V} d\sigma \,,
\label{eq:2_thermo}
\end{equation}
where $\rho$ is the total energy density and $\sigma$ is the entropy per particle. To relax the assumption of an iso-entropic Universe while allowing for matter creation, we adopt the ansatz that the entropy per particle remains constant, i.e.\ $d\sigma = d(S/N)=0$.
This implies that particle creation leads to an overall increase in the system's entropy. Considering an homogeneous and isotropic background, i.e., the FLRW metric in Eq.~\eqref{eq:FLRW_lineele}, it is possible to rewrite Eq.~\eqref{eq:2_thermo} in the form of a generalized continuity equation:

\begin{equation}
\frac{d \rho}{dt} = - 3H (\rho + P) \left( 1 - \frac{1}{3}\frac{d \ln N}{d \ln a} \right)\,,
\label{eq:3_thermo}
\end{equation}
where we set the fiducial volume as $V(t) = a(t)^3 V_0$ and $V_0$ is a fiducial coordinate volume which does not influence the physics content of dynamics. 

A robust phenomenological conjecture associates the process of matter creation with the expansion rate of the Universe \cite{matcre_calvaoLima, matcre_montani2001, elizalde_odintsov}. Accordingly, we consider the following expression:

\begin{equation}
\frac{d \ln N}{d \ln V} = \left( \frac{H}{H_c} \right)^{2 \beta} \,,
\label{eq: ansatz_creation}
\end{equation}
where $H_c$ is a constant quantity, and $\beta > 0$ is a free parameter of the model. \\
By substituting this specific ansatz, we rewrite the generalized continuity Eq.~\eqref{eq:3_thermo}, which accounts for particle creation in the Universe:

\begin{equation}
\frac{d \rho}{dt} = -3 H (\rho + P) \left[1 - \left( \frac{H}{H_c} \right)^{2\beta}\right] \,.
\label{eq:mod_conti_eq}
\end{equation}
To implement matter creation and mimic a reheating phase within our inflationary framework, we identify the newly created particles as a radiation component. In this scenario, the creation mechanism leaves the intrinsic nature of particles unchanged and affects only their number density. Thus, considering a radiation-like component, we adopt the equation of state $P_r = \frac{1}{3} \rho_r$, so that the Eq. \eqref{eq:mod_conti_eq} becomes:

\begin{equation}
\frac{d \rho_r}{dz} = \frac{4}{1 + z} \rho_r \left[ 1 - \left( \frac{H}{H_c} \right)^{2\beta} \right]\,,
\label{eq:mod_continuity_rad}
\end{equation}
where we introduce the redshift variable $z(t) = \frac{a_0}{a(t)} - 1$, such that time differentiation follows the relation $\frac{d}{dt} = -(1 + z) H \frac{d}{dz}$. 

It is worth emphasizing a key aspect of the proposed reheating mechanism, which differs significantly from those discussed in \cite{Appleby:2009uf, Motohashi:2012tt, Dorsch:2024nan, Vilenkin:1985md}. 
In contrast to the quantum field theory (QFT) approach based on vacuum polarization effects, on which the analyzes of \cite{Parker:1968mv, Zeldovich:1971mw, Ford:1986sy} are built upon (see also the reviews \cite{Ford:2021syk, Kolb:2023ydq}), we describe the process of matter creation within a thermodynamic framework. In this picture, the Universe is treated as an open thermodynamic system, in which the number of particles within each comoving volume element can increase, leading to a growth of entropy \cite{matcre_calvaoLima, matcre_montani2001}.
The gravitational origin of matter creation is encoded in the assumed ansatz (Eq.~\ref{eq: ansatz_creation}), which relates the rate of particle production to the time dependence of the Hubble parameter.

In this context, it is important to clarify that the particle creation mechanism considered here refers to the production of elementary constituents of the cosmological fluid, which are by definition chargeless and spinless. 
In particular, we considered the case of a radiation component that at microscopic level corresponds to a thermal bath of elementary particles with different spin: photons have spin $1$, while ultra-relativistic fermions carry spin $1/2$. Therefore, the matter creation mechanism proposed here should be interpreted as a phenomenological description of the average effect over each cosmological fluid element of the microscopic particle production processes, allowed by quantum field theory in an isotropically expanding Universe. In this sense, our approach provides a macroscopic representation of the various sources of particle creation that may operate in a rapidly expanding early Universe.

Furthermore, in the proposed framework, the matter creation process itself is responsible for reheating the Universe. By generating a radiation component, we increase the otherwise low temperature characteristic of a de Sitter phase. 
Unlike models in which reheating occurs after the de Sitter stage through scalar field dissipation (see, e.g., \cite{Motohashi:2012tt, Dorsch:2024nan, Appleby:2009uf}), in our scenario reheating takes place during the de Sitter phase itself. In this picture, the cooling of the Universe due to exponential expansion is compensated by continuous particle production, driven by the large expansion rate of the early Universe.

\subsection{Dynamical formulation of the theory\label{sec: formulation}}

In this work, we study $f(R)$ modified gravity combined with particle creation to build a dynamical framework capable of producing a de Sitter–like inflationary phase, which we then match to the radiation-dominated era of the standard $\Lambda$CDM model as our post-inflation boundary condition. Our approach consists in formulating the resulting dynamical system under suitable assumptions and analyzing the existence, stability, and cosmological viability of its solutions.

The model is developed starting from Eqs.~\eqref{eq:Friedmann_mod} and \eqref{eq:vincolo_FLRW}, considering the modified continuity equation for radiation in Eq.~\eqref{eq:mod_continuity_rad} and assuming that the potential takes the following form:

\begin{equation}
V=2\chi \rho_{\Lambda} + U \,,
\label{eq:ipotesi1}
\end{equation}
where $\rho_{\Lambda}$ 
denotes a constant energy density contribution\footnote{We emphasize that $\rho_{\Lambda}$ indicates the part of the energy of the scalar field $\phi$ associated with the potential, and we choose to indicate it with $\Lambda$ because it represents the primordial cosmological constant term which, in the standard model, induces an inflationary phase.}, and $U(\phi)$ is a generic function. In this way, we are implementing a deviation of the modified gravity from standard de Sitter formulation. 
By implementing this condition in Eq.~\eqref{eq:ipotesi1} and explicitly considering the energy-density contributions, Eqs.~\eqref{eq:Friedmann_mod}, \eqref{eq:vincolo_FLRW} for the dynamics and the modified continuity equation (see Eq.~\eqref{eq:mod_continuity_rad}) can be rewritten here as:
\begin{subequations}
\begin{equation}
H^{2}=\frac{\chi}{3\phi} \left(\rho_r+\rho_{\Lambda}\right)+H^2\frac{\phi'}{\phi}+\frac{U}{6\phi} \,,
\label{eq:dinamica_2}
\end{equation}

\begin{equation}
U' = 6H \phi'\left(2H-H'\right)\,,
\label{eq:vincolo_2}
\end{equation}

\begin{equation}
\rho_r'=4\rho_r \left(1-\left(\frac{H}{H_C}\right)^{2\beta}\right)\,,
\label{eq:continuità_2}
\end{equation}
\end{subequations}
respectively. Here, we change the time variable to $x(z)=\ln(1+z)$ such that $d/dt= -(1 + z) H \,d/dz=-H(x)\, d/dx \equiv - H(x) (\ldots)^{\prime}$. 
It is worth noting that the variable $x$ corresponds to the e-folding number $\mathcal{N}$ commonly used in inflationary scenarios, more precisely it follows the rule $\mathcal{N}=x_{in}-x_{end}$, where $x_{in}$ denotes the epoch at which the slow-rolling phase starts, and $x_{end}$ denotes the epoch at which the inflationary phase ends.

In order to obtain a solvable system for the dynamics (for similar works see \cite{montani_mary_fR, matcre_montanimary}), we impose an additional condition that decouples the Friedmann equation \eqref{eq:dinamica_2} into the following two equations\footnote{We discuss in \autoref{sec: appB}, the other possible choice and why it is not a viable model.}:

\begin{subequations}
\label{eq:disaccoppiamento}
\begin{equation}
H^2= \frac{U}{6\phi}{}\,,
\label{eq:sm_1}
\end{equation}
\noindent \text{and}
\begin{equation}
H^2=-\frac{\chi}{3\phi'} \left(\rho_r+\rho_{\Lambda}\right) \,.
\label{eq:sm_2}
\end{equation}
\end{subequations}

\noindent The condition above selects, among all the possible solutions, the one that ensures to remain as close as possible to the $\Lambda$CDM-model, apart from the scaling for the factor $-1/\phi'$ (\emph{de facto} an effective rescaling of the Einstein constant emerges only via this additional condition). Actually, we are adding a new dynamical equation to our dynamical system, but this is allowed because we promote the function $V(\phi (x))\equiv V(x)$ to be one of the evolution system unknowns. In practice, we reconstruct the potential $V(\phi)$ only \emph{a posteriori}. 
\noindent Hence Eqs. \eqref{eq:sm_1}-\eqref{eq:sm_2} together with Eq.~\eqref{eq:vincolo_2}, provide a system of three equations for the three unknowns $\phi(x)$, $H(x)$ and $U(x)$. \\
Finally, combining Eqs. \eqref{eq:sm_1}-\eqref{eq:sm_2}, we can get a differential equation for the scalar field 
\begin{equation}
     \phi '= -\frac{\chi}{3} \left(\rho_r+\rho_{\Lambda}\right)\frac{6\phi}{U} \,,
    \label{eq:phi_prime}
\end{equation}
and by dividing Eq.~\eqref{eq:vincolo_2} by $H^2$, substituting Eqs.~\eqref{eq:sm_1},\eqref{eq:phi_prime}, we obtain the equation:
\begin{equation}
    U'=-\frac{4U-2\chi (\rho_r+\rho_{\Lambda})}{2U-2\chi (\rho_r+\rho_{\Lambda})}2\chi (\rho_r+\rho_{\Lambda})\,,
    \label{eq:U_prime}
\end{equation}

Now, in order to normalize the cosmological dynamical equations, we define
\begin{equation}
    H^{*^2}=\frac{\chi}{3} \rho_{\Lambda} \,.
    \label{eq: H_star}
\end{equation}
Consequently, the equations governing the cosmological dynamics, namely Eqs.\eqref{eq:sm_2}, \eqref{eq:phi_prime}, \eqref{eq:U_prime}, \eqref{eq:continuità_2}, transform into:
\begin{subequations}
\begin{equation}
H^2 = -\frac{H^{*^2}}{\phi'} \left( \Omega_{r^*} + 1 \right) \,,
\label{eq:Hubble_mod}
\end{equation}
\begin{equation}
    \phi '= -\left(\Omega_{r^*} + 1\right)\frac{\phi}{U^*} \,,
\end{equation}
\begin{equation}
     U^{*'}=-\frac{4U^*-(\Omega_{r^*}+1)}{2U^*-(\Omega_{r^*}+1)} (\Omega_{r^*}+1) \,,
\end{equation}
\begin{equation}
    \ \Omega_{r^*}' = 4 \Omega_{r^*} \left(1 - \frac{\left(-\frac{\Omega_{r^*} + 1}{\phi'} \right)^{\beta}}{H_{c*}^{2\beta}} \right) \,,
    \label{eq:Omega_r_star}
\end{equation}
\begin{equation}
    V^*=1 + U^*\,,
\end{equation}
\label{eq:sist_finale}
\end{subequations}
\noindent where we introduced the normalized parameters:
\begin{equation}
	\Omega_{r^*}\equiv \frac{\rho _r}{\rho^*}\, ,\,  H_c^* \equiv \frac{H_c}{H^*} \, , \, V^* \equiv \frac{V}{6H^{*^2}}
	\,.
	\label{eq: par_norm}
\end{equation}

\noindent The resulting coupled differential equations for the variables $\phi(x)$, $U^*(x)$ and $\Omega_{r^*}(x)$ can be solved numerically by imposing natural boundary conditions:

\begin{equation}
    \phi(x_{\rm end})=1, \quad U^*(x_{end})=U^*_{\rm end}\quad \mathrm{and} \quad  \Omega_{r^*}(x_{\rm end})=\Omega_{r^*_{end}}\,,
    \label{eq: boundary_cond}
\end{equation} 
where we remind that $x_{\rm end}$ represents the redshift at which the inflationary phase ends and we imposed $\phi(x_{end})=1$ since we require the matching condition with a $\Lambda$CDM model, i.e. with GR.

Hence, the free parameters of the model are classified into two categories:
\begin{enumerate}
    \item Parameters related to the matter creation mechanism: $\beta$, $H_c^*$ and $\Omega_{r^*_{end}}$; 
    \item Parameters associated with the modified gravity dynamics: $x_{\rm end}$, $H^*$ and $U^*_{\rm end}$. 
\end{enumerate}

\section{Model testing} \label{sec: model_testing}
In \autoref{sec: formulation}, we present an $f(R)$ gravity model capable of generating a de Sitter-like inflationary phase, when incorporating particle creation. This phase is characterized by a modified Hubble function Eq.~\eqref{eq:Hubble_mod}, which would leave detectable imprints throughout the entire cosmic history. In this section we test the model presented in the previous sections, using late-time observable that would be affected by this modified Hubble function.

\subsection{Sampling the posterior distribution}
The aim of this work is to test the proposed model against available observational data, both to assess its physical validity and to place constraints on the parameters describing it. To achieve this, we sample the posterior distribution of the model and cosmological parameters using a Bayesian statistical framework \cite{bayesian}.
The reconstruction of the posterior is done using the \texttt{Nautilus} \cite{nautilus} sampler interfaced with \texttt{Cobaya} \cite{Cobaya}. The theoretical modeling of the unmodified late-time quantities is handled using \texttt{CAMB}~\cite{Lewis:1999bs}, while custom modifications to early-time, including the evaluation of $H_0$ that will be described by Eq.~\eqref{eq:H0_der}, are implemented within a dedicated theory module. These modifcations are implmented directly in \texttt{CANDI}\footnote{\href{https://github.com/chiaradeleo1/CANDI}{https://github.com/chiaradeleo1/CANDI}} \cite{CANDI}.
\begin{table*}[htp!]
\centering
\begin{tabularx}{\textwidth}{m{6cm}m{3cm}m{2cm}}
\multicolumn{2}{l}{\textbf{Parameters}}  & \textbf{Prior} \\
\hline
\hline
\multicolumn{3}{c}{\textbf{\quad\quad\quad\quad Cosmological parameters}} \\
\hline
Present day matter content & $\Omega_m$ & $\mathcal{U}(0.1, 0.99)$ \\
Present day baryonic density &$\Omega_b\,h^2$ & $\mathcal{N}(0.005, 0.1)$ \\
Absolute Magnitude &$M_B$ & $\mathcal{U}(-15, -25)$ \\
Hubble constant &$H_0$ & derived \\
Number of effective neutrinos & $N_\nu$ & derived\\
\hline
\hline
\multicolumn{3}{c}{\textbf{\quad\quad\quad\quad Model parameters}} \\
\hline
Logarithm of End of Inflation & $x_{\rm end}$ & $\mathcal{U}(55, 58.5)$ \\
Logarithm of $H^*$ & $\rm log_{10}H^*=\gamma$ & $\mathcal{U}(47.5, 45.5)$ \\
Logarithm of $U^*_{\rm end}$ & $\rm log_{10}(U^*_{\rm end})$ & $\mathcal{U}(6, 3.5)$ \\
\hline
\end{tabularx}
\caption{Prior distributions for the cosmological and model parameters sampled.}

\label{tab:prior}
\end{table*}
\subsection{Model, Data and sampled parameters} \label{sec:model}

To test the impact of the modified inflationary phase on present-time observables, it is necessary to establish a connection between early-time and late-time quantities. Indeed, a modified early-time Hubble function can leave observable imprints in late-time cosmology. Therefore, we rely on late-time probes, specifically Baryon Acoustic Oscillations (BAO) from DESI DR2 \cite{desicollaboration2025desidr2resultsii} and Type Ia Supernovae (SNIa) from the Pantheon$^+$ catalog \cite{Brout_2022}. 
Although we do not perform a full CMB analysis in this work, we verify that the theoretical predictions of our model for the primordial spectral parameters $n_s$ and $r$ are consistent with current CMB constraints. We stress, however, that these observational bounds are derived within a $\Lambda$CDM framework. A fully consistent assessment of our model will require incorporating the complete CMB information, including the scalar perturbation equations, which we leave to future work.

To consistently relate the modified early-time dynamics to late-time observations, we build our model under the assumption that, at a certain epoch after the end of inflation ($x < x_{\text{\rm end}}$), a phase transition occurs, restoring the standard cosmological evolution and initiating the radiation-dominated era. This transition requires the modified Hubble function to smoothly match the standard Hubble function after inflation, as we will show in \autoref{sec:theo_results}. \newline
Starting from Eq.~\eqref{eq:Hubble_mod}, we impose that, at $x = x_{\text{\rm end}}$, the expression for modified $H(x)$ must match the standard form valid during the radiation-dominated epoch. This leads to the condition:
\begin{equation}
    H_0 = \frac{H^*}{E(x_{\rm end})} \sqrt{\frac{\Omega_{r^*_{end}}+ 1}{-\phi^\prime(x_{\rm end})}}, 
\label{eq:H0_der}
\end{equation} 

where $E(x_{\rm end})$ is Hubble rate parameter in $\Lambda$CDM.
We use this expression to test the model by sampling the parameters that describe it, specifically $x_{\rm end}$, $\gamma$, and $U^*_{\rm end}$, along with the standard cosmological parameters. In particular, we aim to identify possible degeneracies among them. The sampled parameters and their associated priors are listed in \autoref{tab:prior}, while the others are kept fixed at the reference values shown in \autoref{tab:fiducial_model_params}. 
\begin{table}[h!]
\centering
\renewcommand{\arraystretch}{1.2}
\begin{tabular}{l|ccccc}
\hline
Parameters & $\beta$ & $H_c^*$ & $\Omega_{r_{\mathrm{end}}}^*$ &  $\phi_{\rm end}$ \\
\hline
Value      & $10^{-3}$ & $0.1$     & $10$ &  $1$   \\
\hline
\end{tabular}
\caption{Fiducial values adopted for the model parameters that are held constant in the analysis. The theoretical justification for these choices is provided in \autoref{sec:appA}.}
\label{tab:fiducial_model_params}
\end{table}
We set our fiducial values assuming that the Universe transitions in a radiation dominated epoch, we select $H^* \sim 10^{46}$, $\Omega_{r_0} \sim 8 \times 10^{-5}$, $U^*_{\rm end}\sim 10^5$ and $z_{\rm end} \sim 10^{24}$ (corresponding to $x_{\rm end} \sim 56$). We fix $\beta = 10^{-3}$, $H_c^* = 0.1$, and $\Omega_{r^*_{\rm end}} = 10$, since these parameters do not affect the background evolution. Therefore, the observables considered in this work are insensitive to them and they cannot be meaningfully constrained within the present sampling analysis. Their values are thus chosen on theoretical grounds. A full perturbation-level analysis, including the computation of scalar and tensor fluctuations and a consistent comparison with the complete CMB temperature and polarization spectra, would instead be sensitive to these parameters. We leave such an extended analysis to future work. For further details and for the theoretical motivation behind the adopted fiducial values, we refer the reader to \autoref{sec:appA}.

The other model parameters are sampled in logarithmic space.
It should be noted that we do not impose a direct prior on $H_0$, since in our framework it is a derived parameter obtained from Eq.~\eqref{eq:H0_der}. However, in order to preserve its physical interpretation as the present-day Hubble constant, we require it to lie within the conservative interval $H_0 \in [20,100]\,\mathrm{km\,s^{-1}\,Mpc^{-1}}$. 
This condition is implemented at the level of the theory module and effectively translates into constraints on the allowed combinations of the sampled parameters.

Since we aim to constrain $H_0$ to ensure a match between the two models at the end of the inflationary phase, the BAO and SNIa data alone are insufficient. Specifically, SNIa alone cannot provide a direct measurement of $H_0$ due to its degeneracy with the absolute magnitude of supernovae, $M_B$. For this reason we include the calibration of $M_B$ from SH0ES Cepheids \cite{Riess_2022}.
Similarly, BAO measurements constrain combinations of parameters such as $H_0\,r_d$ and $\Omega_m$. However, BAO alone cannot disentangle the Hubble parameter $H_0$ from the sound horizon $r_d$. The latter depends directly on the physical baryon density $\Omega_b h^2$ and on the effective number of relativistic species $N_{\nu}$. Therefore, additional information on these quantities is necessary to break the degeneracy between $r_d$ and $H_0$. In our analysis, we follow the DESI approach \cite{desicollaboration2025desidr2resultsii} by adopting an external constraint on the physical baryon density,
\begin{equation}
    \Omega_b h^2 = 0.02196 \pm 0.00063,
\end{equation}
while leaving the effective number of relativistic species $N_{\nu}$ free. We do not fix $N_{\nu}$ a priori, since in our framework it can be directly connected to the physics of the end of inflation.
Indeed, we assume that after the restoration of the standard $\Lambda$CDM evolution the Universe enters a radiation-dominated epoch. In this regime, the radiation density is related to the matter density through
\begin{equation}
    \Omega_r = \frac{\Omega_m}{1+z_{\rm eq}},
    \label{eq:omega_r}
\end{equation}
where $z_{\rm eq}$ denotes the redshift of matter–radiation equality. 
The equality redshift can in turn be related to the end of inflation by using the standard scaling valid during radiation domination,
\begin{equation}
    1+z_{\rm eq} = \sqrt{\frac{t_{\rm end}}{t_{\rm eq}}} (1+z_{\rm end}),
    \label{eq:z_eq_z_end}
\end{equation}
where we take $t_{\rm end} \sim 10^{-32}\,\mathrm{s}$ and $t_{\rm eq} \sim 10^{11}\,\mathrm{s}$ \cite{KolbTurner}. 
Furthermore, following \cite{Zhao_2017}, the effective number of relativistic species can be written as
\begin{equation}
    N_{\nu} = \frac{1}{0.2271} \frac{\Omega_m h^2}{\Omega_\gamma h^2 (1+z_{\rm eq})} - 1,
    \label{eq:neutrinos}
\end{equation}
which, combined with Eq.~\eqref{eq:z_eq_z_end}, establishes a direct link between $N_{\nu}$ and the inflationary parameter $x_{\rm end}$.
Finally, we assume the same covariance between $N_{\nu}$ and $\Omega_b h^2$ adopted in \cite{BBN}.\newline

\section{Parameter Estimation Results}\label{sec:results}
Considering the parameters listed in \autoref{tab:prior} and the observables discussed in \autoref{sec: model_testing}, we present the results on the model parameters and the cosmological one. All the results for the different cases and dataset can be read in \autoref{tab:results}. 
\begin{figure}
    \centering
    \includegraphics[width=1\linewidth]{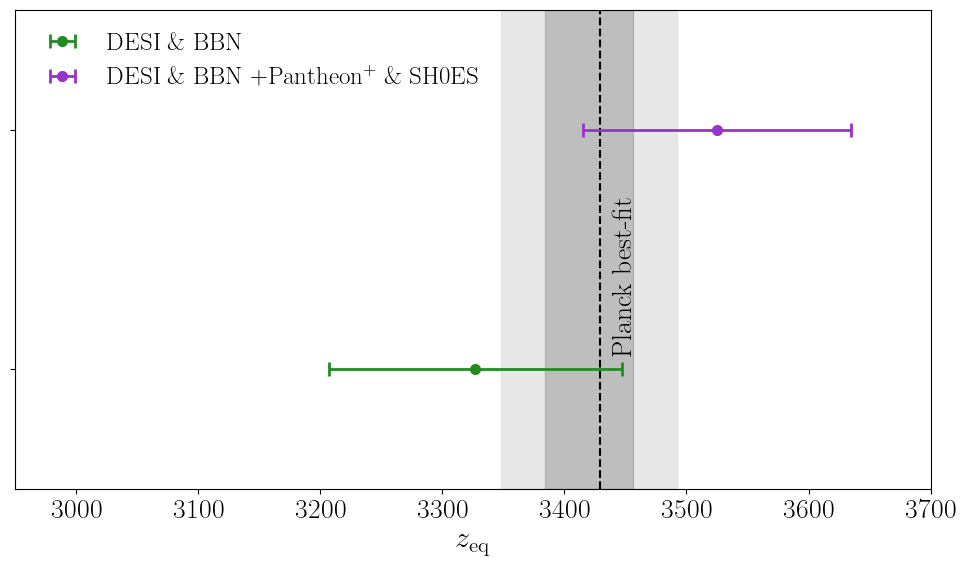}
    \caption{$1\sigma$ constraint on $z_{\rm eq}$ obtained using the DESI+BBN dataset (green) and the full one using DESI+BBN+Pantheon$^+$+SH0ES (purple). The black dashed line represent the Planck18 best-fit \cite{Planck18}, while the gray areas are the $1\sigma$ confidence region (more opaque) and the $2\sigma$ region (more transparent).}
    \label{fig:z_eq}
\end{figure}

As mentioned in \autoref{sec: model_testing}, since the neutrinos are allowed to vary freely, we derive the equivalence epoch using Eq.~\eqref{eq:z_eq_z_end}. In \autoref{fig:z_eq}, we show that the value obtained for the equivalence epoch is compatible within $1\sigma$ with the Planck18 best-fit results \cite{Planck18}, represented by the black dashed line, while the gray area indicates the $1\sigma$ confidence region (more opaque) and the $2\sigma$ region (more transparent). We show the $1\sigma$ bound from the joint analysis of the DESI dataset with a BBN prior and the Pantheon$^+$ dataset with the SH0ES calibration (purple) and the one obtained with DESI and BBN only (green).

\begin{figure}[h!]
    \centering
    \includegraphics[width=1.\linewidth]{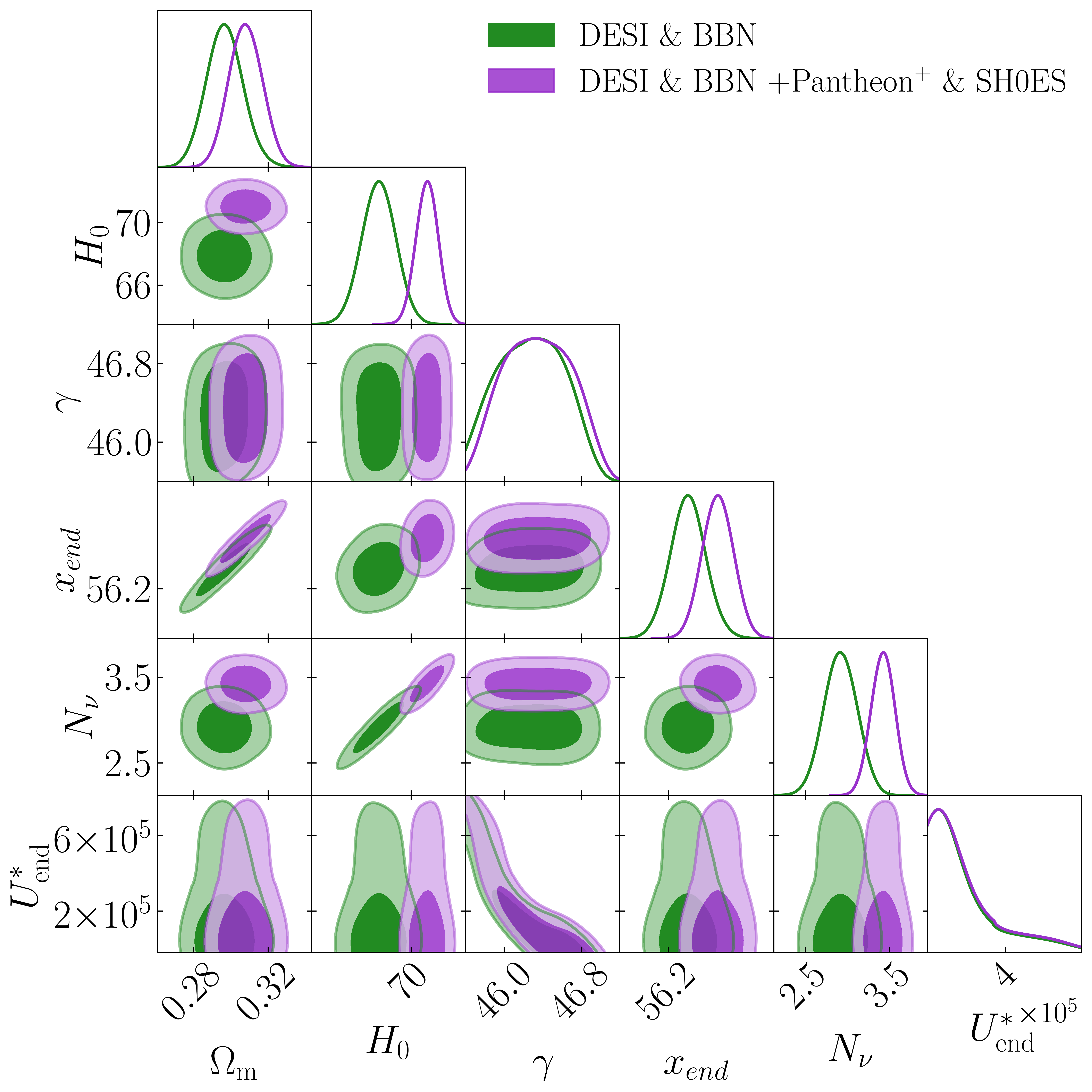}
    \caption{One and two dimensional marginalized constraints on the cosmological and model parameters listed in \autoref{tab:prior}, comparing the DESI+BBN dataset (green) with the full dataset (purple). }
    \label{fig:triangle}
\end{figure}

In \autoref{fig:triangle}, we present the results for the cosmological and model parameters listed in \autoref{tab:prior} using DESI+BBN and DESI+BBN combined with Pantheon$^+$+SH0ES. The other model parameters not shown in the plots are kept fixed at the value shown in \autoref{tab:fiducial_model_params}.
Overall, the constraints on the cosmological paameters remain consistent with DESI+BBN, and notably, we recover a higher value of $H_0$ when the SH0ES calibration is included. It is important to note that the analysis performed with Pantheon$^+$+SH0ES dataset alone is not shown in the plot, as it is not sensitive to variations in neutrinos or the end of inflation; instead, it serves primarily to break parameter degeneracies.
Regarding the model parameters, we find a clear degeneracy between the present-day matter density parameter $\Omega_m$ and the end of inflation $x_{\rm end}$. This behaviour is expected from Eqs.~\eqref{eq:omega_r} and \eqref{eq:z_eq_z_end}, where both quantities enter the relation between the inflationary epoch and the subsequent radiation- and matter-dominated expansion. In particular, larger values of $\Omega_m$ are compatible with an inflationary phase ending at earlier times (larger $x_{\rm end}$). Physically, a higher matter content today implies a longer effective expansion history after inflation, allowing the Universe to reach the same observed late-time state even if inflation terminates earlier.

Looking at the remaining model parameters, we observe a clear degeneracy between $\gamma$ and $U^*_{\rm end}$. This behaviour can be understood by noting that both parameters affect the overall amplitude of the Hubble expansion rate. As shown in Eq.~\eqref{eq:sm_1}, the Hubble parameter is directly related to the effective potential $U$, establishing a direct link between the normalized potential evaluated at $x_{\rm end}$, $U^*_{\rm end}$, and the amplitude of $H^2$.
In the modified formulation, this dependence is encoded implicitly in the background evolution equations. In particular, Eqs.~\eqref{eq:Hubble_mod}-\eqref{eq:Omega_r_star} show that $H^2(x)$ directly depends on $H^*$, whose magnitude is controlled by the exponent $\gamma$, and the effective quantity $U^*(x)$, whose normalization is fixed by the boundary condition $U^*_{\rm end}$. As a consequence, variations in $\gamma$ can be compensated by opposite variations in $U^*_{\rm end}$, leaving the background expansion history $H(z)$ nearly unchanged.

We need to carefully analyze the behavior of the parameter $U^*_{\rm end}$. 
From the triangle plot in \autoref{fig:triangle}, the marginalized posterior in linear space might appear to be cut off at low values of $U^*_{\rm end}$, which could suggest that the prior range should be extended.  If this were the case, one would expect the sampled points to accumulate close to that apparent boundary. 
However, this is not observed: the samples are instead uniformly distributed when inspected in terms of $\log_{10}(U^*_{\rm end})$.
This behavior can be understood by noting that we assume a flat prior on $\log_{10}(U^*_{\rm end})$. 
As a consequence, $\log_{10}(U^*_{\rm end})$ is largely unconstrained by the data and its posterior distribution is dominated by the assumed prior. 
When transforming the variable from $\log_{10}(U^*_{\rm end})$ to $U^*_{\rm end}$, the change of variables introduces a Jacobian factor, resulting in a posterior distribution proportional to $1/U^*_{\rm end}$. 
The apparent decrease observed in the triangle plot therefore reflects this transformation and does not indicate a preference driven by the data.
Nevertheless, \texttt{getdist} reports an effective constraint,
$U^*_{\rm end} = 1.4^{+0.4}_{-2}\times 10^{5},$
which should be interpreted with care. 
This value is primarily driven by the prior volume and by the additional requirement that the background evolution yields a physically acceptable value of the Hubble constant $H_0$. 
In particular, the hard prior imposed on $H_0$ excludes part of the $U^*_{\rm end}$ parameter space, effectively acting as an upper bound rather than a genuine observational constraint.

\begin{figure}[h!]
     \centering
     \includegraphics[width=\linewidth]{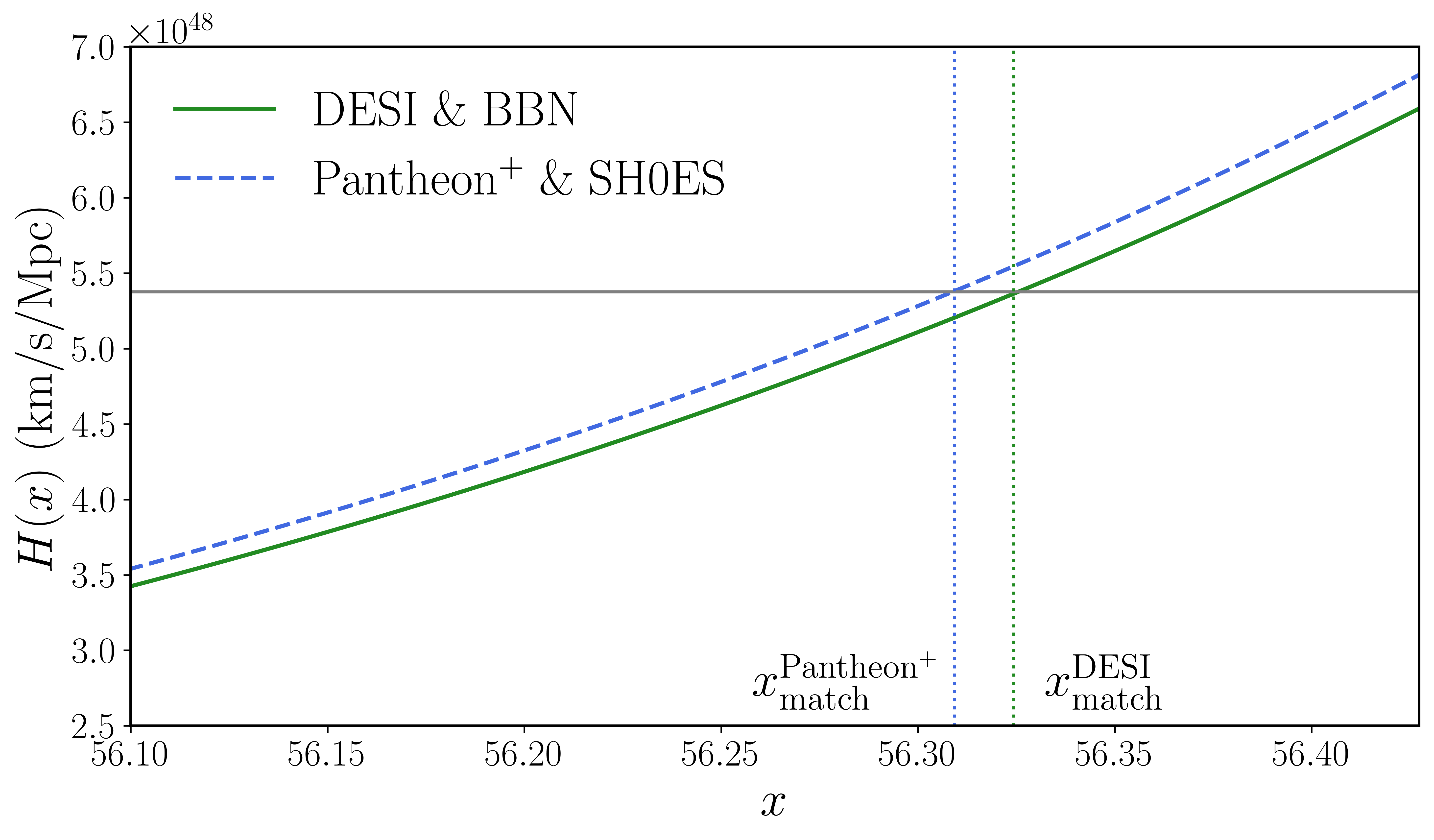}

    \caption{
Reconstructed Hubble function $H(x)$ for the DESI+BBN (green) and Pantheon$^+$+SH0ES (blue) datasets.
The horizontal dashed line shows the value of $H(x_{\rm end})$ computed in $\Lambda$CDM with \texttt{CAMB}, and each curve’s intersection with this line marks the inferred end of inflation epoch required to reproduce the observed $H_0$.}

    \label{fig:Hx}
\end{figure}

In \autoref{fig:Hx}, we reconstruct the Hubble function for the DESI+BBN and Pantheon$^+$+SH0ES datasets. Specifically, we start from the independent constraints on $H_0$ and $\Omega_r$ obtained from each dataset, and we use the values to compute $H(x)$ using the standard expression. Then, we evaluate the value of $H(x)$ at the end of inflation within the $\Lambda$CDM framework using \texttt{CAMB}, which is represented by the gray horizontal line in the figure. We identify the end of inflation as the epoch when the reconstructed $H(x)$ curve intersects this horizontal line. This crossing point can be interpreted as the moment when inflation, as described by our model, should end in order to reproduce the observed present-day value of $H_0$.
As shown, the tension in $H_0$ measurements translates into a corresponding tension in determining the end of inflation.

\begin{table*}[h!]
\centering
\renewcommand{\arraystretch}{1.6}
\setlength{\tabcolsep}{12pt}

\resizebox{\textwidth}{!}{
\begin{tabular}{
>{\centering\arraybackslash}p{3.0cm} |
c c c c c c}

    \hline
     & \multicolumn{6}{c}{\textbf{Sampled parameters}} \\
    \cline{2-7}
    \raisebox{2.5ex}{\textbf{Data}}
     & $H_0$ & $\Omega_m$ & $x_{\rm end}$ & $\gamma$ & $U^*_{\rm end}$ & $N_{\nu}$ \\
    \hline
    \hline

    \parbox[c][3.4cm][c]{3.0cm}{\centering\rotatebox{90}{DESI + $\rm Pantheon^+$}}
    & $71.04 \pm 0.71$
    & $0.3079 \pm 0.0088$
    & $56.302 \pm 0.031$
    & $46.36 \pm 0.34$
    & $1.5 ^{+0.4}_{-2} \cdot10^5$
    & $3.42 \pm 0.14$ \\
    \hline

    \parbox[c][2.4cm][c]{3.0cm}{\centering\rotatebox{90}{DESI}}
    & $67.9 \pm 1.1$
    & $0.2966 \pm 0.0096$
    & $56.241 \pm 0.036$
    & $48.30 \pm 0.33$
    & $1.4 ^{+0.4}_{-2} \cdot10^5$
    & $2.92 \pm 0.19$ \\
    \hline

    \parbox[c][2.4cm][c]{3.0cm}{\centering\rotatebox{90}{$\rm Pantheon^+$}}
    & $73.4^{+1.1}_{-1.0}$
    & $0.336 \pm 0.017$
    & $56.12 \pm 0.53$
    & $46.31^{+0.33}_{-0.54}$
    & $1.2 ^{+0.4}_{-2} \cdot10^5$
    & $8.3^{+4.8}_{-8.9}$ \\
    \hline

\end{tabular}
}

\caption{Results for the cosmological and model parameters using DESI $\&$ BBN and Pantheon$^+$ $\&$ SH0ES datasets and the combined dataset.}
\label{tab:results}
\end{table*}

Lastly, looking at \autoref{tab:results}, we can also draw conclusions regarding the Hubble tension. When the DESI and Pantheon$^+$ data are combined, the tension on $H_0$ relative to the SH0ES measurement \cite{Riess_2022} is reduced to $1.59\sigma$. When the datasets are considered separately, the model yields a $3.7\sigma$ tension between DESI and Pantheon$^+$.

\section{Theoretical results}\label{sec:theo_results}
In this section we reconstruct the theoretical functions characterizing the dynamics of the model using the best-fit results obtained from the data analysis \autoref{sec:results} and summarized in \autoref{tab:bestfit}. We discuss the emergent reheating temperature and we also report the resulting spectral observables, comparing the theoretical predictions with the most recent experimental values.  \\
In particular, we present here the behavior of $H(x)$ and $\phi(x)$ reconstructed by solving the system of coupled differential equations in Eq.~\eqref{eq:sist_finale} and using the best-fit parameters reported in \autoref{tab:bestfit}. 

\begin{figure}[h!]
    \centering
        \centering
        \includegraphics[width=0.8\linewidth]{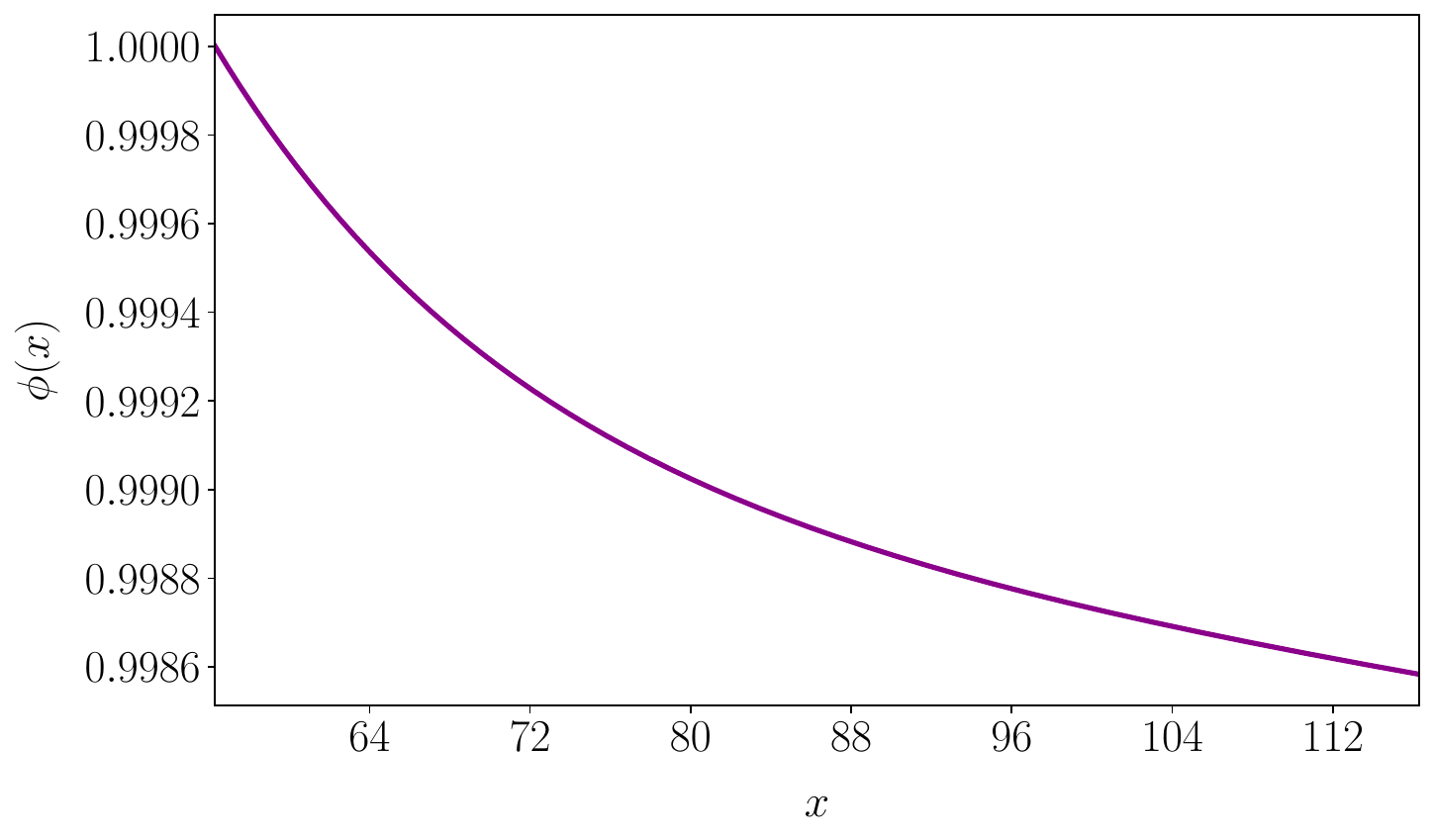}
    \caption{Evolution of the scalar field $\phi(x)$ using the best-fit values in \autoref{tab:bestfit}.}
    \label{fig:phi_theo}
\end{figure}

\begin{figure}[h!]
    \centering
        \centering
        \includegraphics[width=0.75\linewidth]{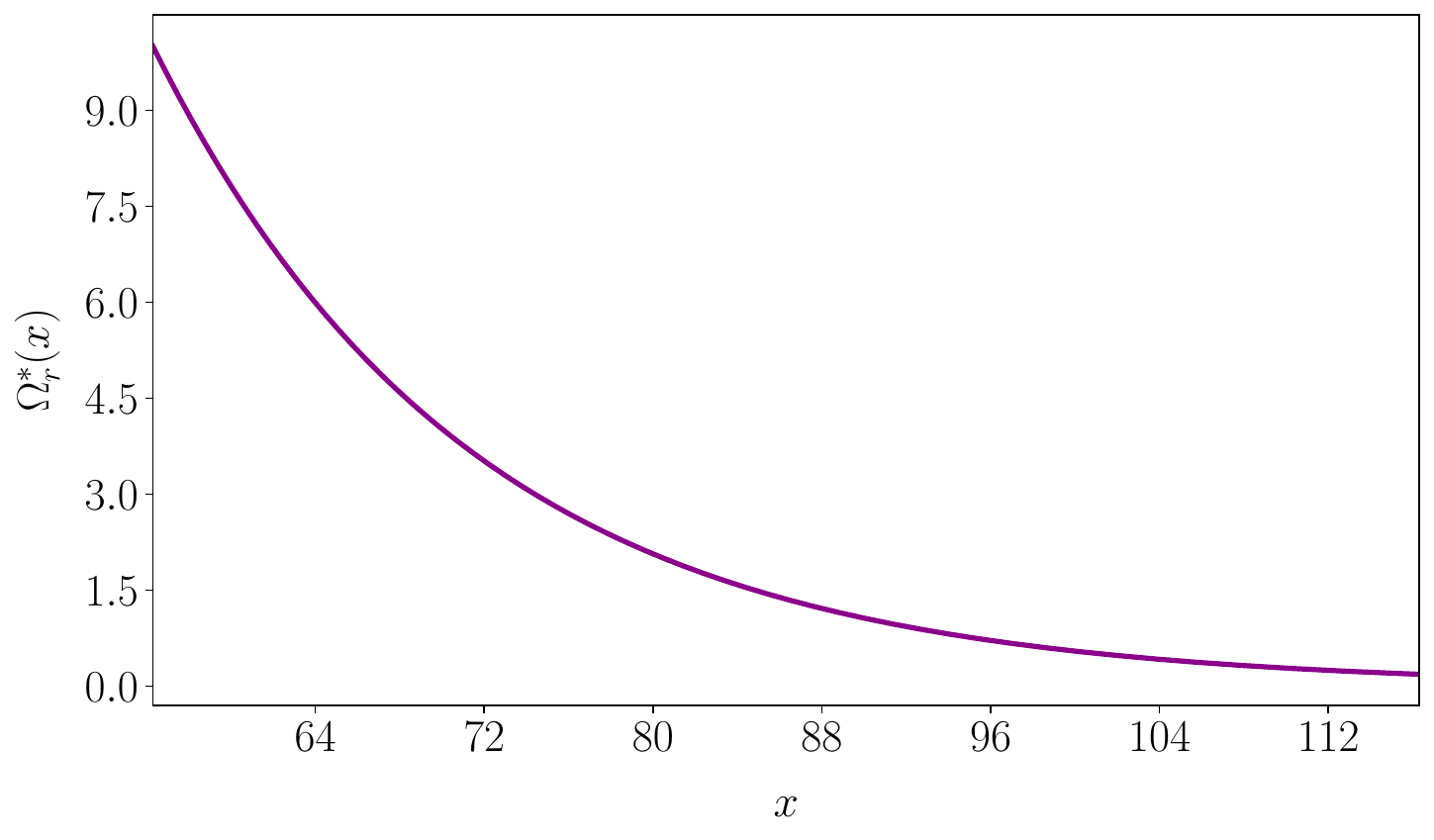}
    \caption{Evolution of $\Omega_r^*(x)=\frac{\rho_r(x)}{\rho_{\Lambda}}$ using the best-fit values in \autoref{tab:bestfit}.}
    \label{fig:omegar_theo}
\end{figure}

As explained in \autoref{sec:model}, we construct a modified gravity inflationary model in which a phase transition at a specific epoch after the end of inflation restores the $\Lambda$CDM cosmology. Before this epoch, the evolution is de Sitter–like, characterized by an approximately constant expansion rate. 
\newline
In \autoref{fig:phi_theo} we show the reconstructed profile of the scalar field $\phi(x)$, which is, as expected, a decreasing function of $x$ starting from $\phi_{end}=1$, representing the $\Lambda$CDM limit at the end of the inflationary period.
Overall, this model successfully reproduces a de Sitter-like phase characterized by an approximately constant Hubble parameter due to the modified dynamics introduced by the $f(R)$ gravity. In this scenario, the scalar field non-minimally coupled to gravity remains nearly constant throughout inflation, resulting in a slow-rolling behavior. 
In \autoref{fig:omegar_theo} we can visualize the effect of particle creation mechanism, i.e. it reverses the abundance between the radiation energy density $\rho_r$ and the cosmological constant energy density $\rho_{\Lambda}$. For higher values of $x$, i.e. at the beginning of the inflationary process, radiation is subdominant until $x\sim 100$, when these two components become comparable and radiation-type particles are then overproduced.

\begin{table}[h!]
\centering
\renewcommand{\arraystretch}{1.2}
\begin{tabular}{cccccc}
\hline
$\Omega_m$ & $x_\text{\rm end}$ &  $U^*_{\rm end}$ & $\gamma$  & $H_0$ &$N_\nu$\\
\hline
 0.3079 & 56.30247 & $10^{4.74}$ & 46.3565 & 71.036  & 3.4227 \\
\hline
\end{tabular}
\caption{Best-fit cosmological parameters for DESI $\&$ BBN and Pantheon$^+$ $\&$ SH0ES combination obtained from the sampling analysis.}
\label{tab:bestfit}
\end{table}

Furthermore, based on the results from the model testing, we compute the final reheating temperature based on the Hubble parameter at the matching point. It is easy to check that the final temperature of our reheating process is approximately $10^{10}$ GeV, which is clearly in agreement with all the observational constraints \cite{weinberg, KolbTurner}.

\subsection{Predictions on primordial power spectrum observables}
In modified-gravity inflation, quantized perturbations of the 
scalar field $\phi$ in the Jordan frame generate the primordial power spectrum. The resulting predictions for 
$n_s$ and $r$ can be compared with observations to assess the model’s viability.
We computed the inflationary observables $n_s$ and $r$ predicted by our theory using the best-fit values from the background analysis reported in \autoref{tab:bestfit}, in order to assess, at a preliminary stage, if the model is compatible with the most recent constraints on the same observables. \\
Given an $f(R)$-inflationary theory in the Jordan frame, it is possible to define the following variables \cite{Hwang:1996xh, Hwang:2001pu, DeFelice:2010aj, Ivanov:2021chn, Ivanov:2025nsx}:
\begin{equation}
    \epsilon_1=\frac{H'}{H}, \quad \epsilon_2=-\frac{\epsilon_1 '}{\epsilon_1} \,,
\end{equation}
and 
\begin{equation}
    \epsilon_3=\frac{\phi'}{2\phi}, \quad \epsilon_4=-\frac{\epsilon_{3}'}{\epsilon_{3}} \,.
\end{equation}
In particular, after verifying that at the epoch where the primordial power spectrum is evaluated the radiation component is subdominant with respect to the scalar field (see \autoref{fig:omegar_theo}), and that all the relevant spectral parameters satisfy $\epsilon_i\ll 1$, we can safely adopt the following expressions for the scalar spectral index and the tensor-to-scalar ratio:
\begin{equation}
    n_s\simeq1+6\epsilon_3-2\epsilon_4\,,
    \label{eq:ns_theo}
\end{equation}
\begin{equation}
    r\simeq\epsilon_4^2.
    \label{eq:r_theo}
\end{equation}

\begin{figure}[h!]
    \centering
    \begin{subfigure}[b]{0.8\linewidth}
        \centering
        \includegraphics[width=\linewidth]{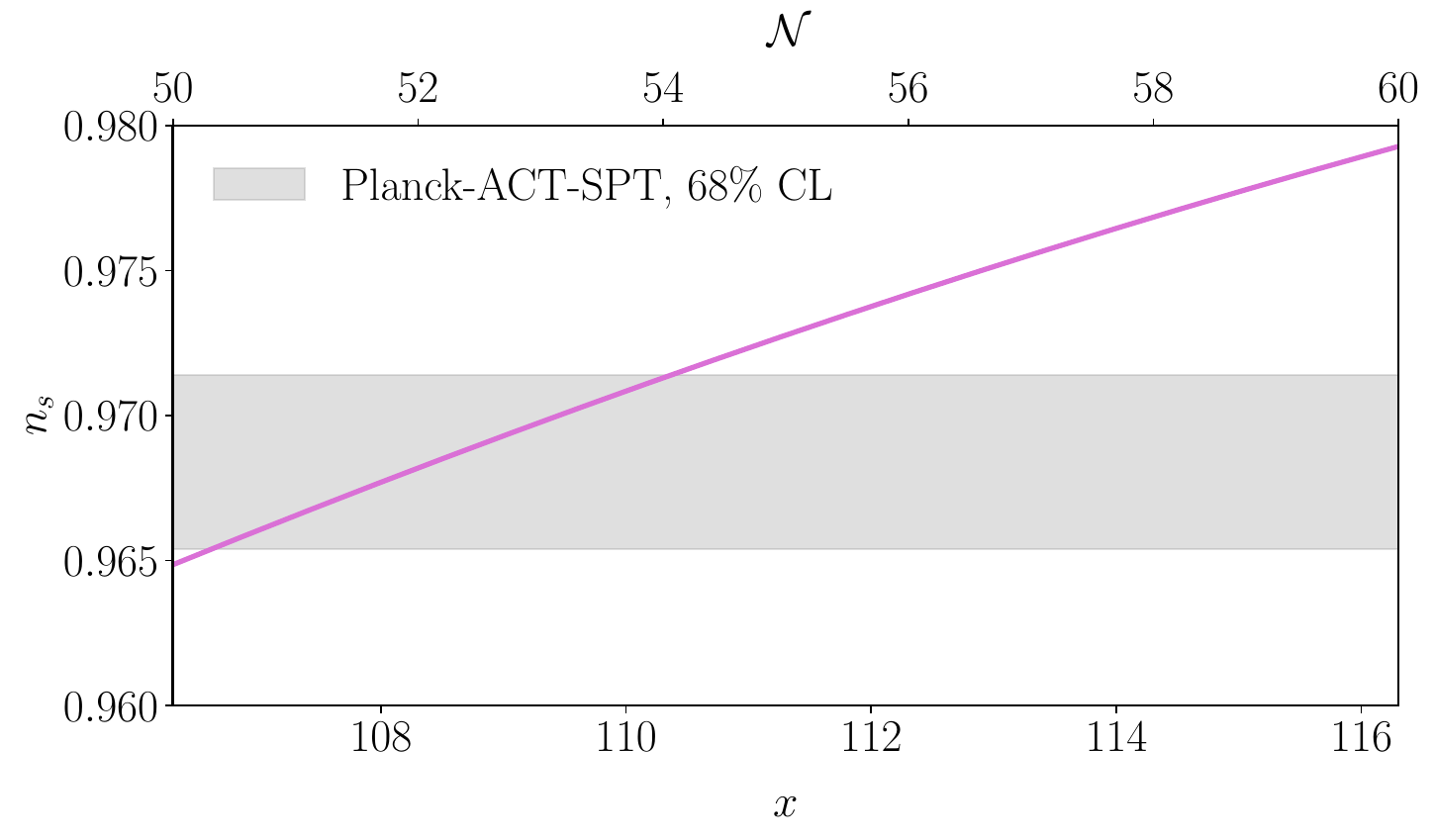}
        \caption{}
        \label{fig:ns}
    \end{subfigure}
    \hfill
    \begin{subfigure}[b]{0.8\linewidth}
        \centering
        \includegraphics[width=\linewidth]{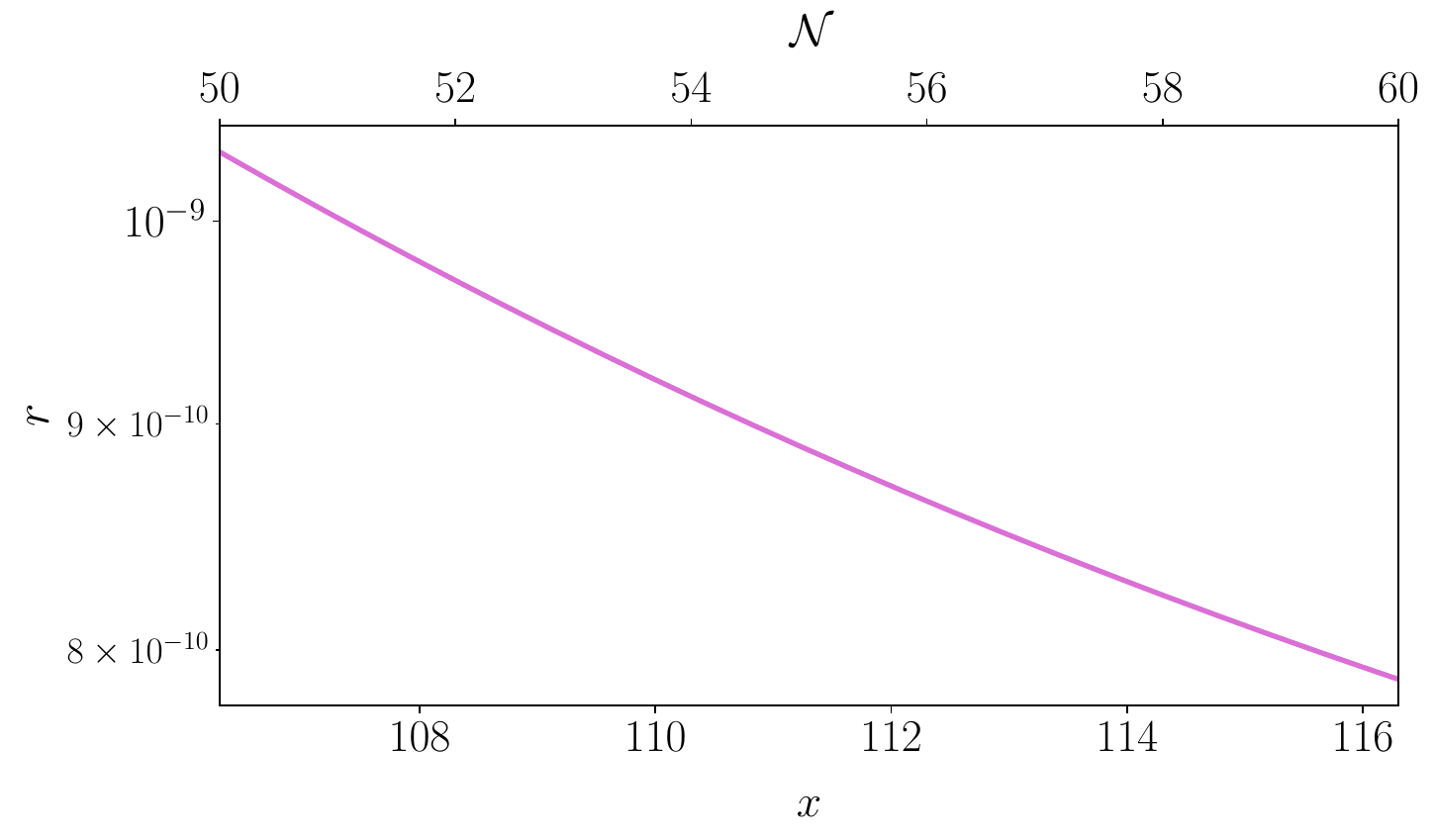}
        \caption{}
        \label{fig:r}
    \end{subfigure}
    \caption{a) Predicted spectral index computed using the best-fit values in \autoref{tab:bestfit}. The gray band represents the $68\%$ C.L. observational constraint on $n_s$ from the most recent combined CMB data from Planck+ACT+SPT \cite{SPT-3G:2025bzu}. b) Predicted tensor-to-scalar ratio obtained with the same best-fit parameters. }
    \label{fig:spectral_obs}
\end{figure}

We show in \autoref{fig:spectral_obs} the predicted spectral observables using the best-fit values and considering a range of \emph{e-foldings} $(50-60)$, which is the standard interval adopted in literature \cite{Liddle:2003as, German:2022sjd, DiMarco:2024yzn, Remmen:2014mia}. In \autoref{fig:ns} the predicted value of the spectral index is compared with the most recent observational value from Planck+ACT+SPT $n_s=(0.9684 \pm 0.003)$ \cite{SPT-3G:2025bzu}, showing compatibility if the perturbations cross the Hubble horizon around $50-54$ e-foldings before the end of inflation. The tensor-to-scalar ratio $r$ in \autoref{fig:r} is in our scenario of order $10^{-9}$, meaning that it satisfies the most stringent upper limit $r \le 0.032$ from BICEP+Keck \cite{BICEP:2021xfz, Tristram:2021tvh}. Overall, this model is compatible with the most recent constraints from CMB experiments.

However, a fully consistent assessment requires the inclusion of the complete CMB dataset and a dedicated treatment of cosmological perturbations within our modified-gravity framework. This more comprehensive analysis, implementing the perturbation equations directly in the model and confronting them with the full CMB likelihood, will be the subject of future work.

\section{Conclusion} \label{sec:conclusion}
In this work, we have performed a background‐level analysis of an $f(R)$ inflationary model in which the modified dynamics sustains a de Sitter-like phase and the particle creation supplies the re-heating phase. The model is defined by a modified Hubble parameter $H(x)$ up to the epoch $x_{\rm end}$, identifying the end of inflation, after which $H(x)$ transitions smoothly to its standard $\Lambda$CDM form. Such an early‐time deformation of the expansion history imprints subtle but measurable effects on late‐time observables.

We compared the model against current data, namely the Pantheon$^+$ catalougue calibrated with SH0ES and the DESI-DR2 BAO combined with BBN priors. Our sampling analysis was carried out varying both cosmological ($\Omega_m$, $\Omega_bh^2$, $N_\nu$) and model ($\gamma$, $x_{\rm end}$, $U^*_{\rm end}$). Letting the number of neutrinos free, we accurately recovered the matter–radiation equality epoch, related to $N_{\nu}$ and $x_{\rm end}$ as detailed in \autoref{sec: model_testing}, in agreement with Planck 2018.We find a degeneracy between the matter density parameter $\Omega_m$ and the end of inflation $x_{\rm end}$. This is consistent with the expectation that an earlier termination of inflation requires a larger present-day matter content. If inflation ends earlier, cosmological perturbations have less time to grow during the inflationary phase; consequently, reproducing the observed late-time Universe demands a higher value of $\Omega_m$. 
\noindent Also, we identify a degeneracy between the parameters $\gamma$ and $U^*_{\rm end}$. This arises because both quantities control the overall normalization of the Hubble expansion rate, entering the background evolution in a correlated manner. As a result, changes in $\gamma$ can be compensated by corresponding variations in $U^*_{\rm end}$, leaving the predicted expansion history $H(z)$ effectively unchanged.
We then reconstruct the Hubble function separately for DESI+BBN and Pantheon$^+$+SH0ES best fits, we demonstrated that the tension in $H_0$ can equivalently be expressed as a tension in the inferred value of $x_{\rm end}$.
Looking at the proper Hubble tension, we see that the combined analysis of DESI and Pantheon$^+$ data reduces the tension with the SH0ES result to $2.81\,\sigma$. Furthermore, considering the tension in $H_0$ between DESI and Pantheon$^+$ themselves, we find that it amounts to $5\,\sigma$.

Using the best-fit parameters obtained from the background analysis, we reconstructed the theoretical functions underlying the modified gravity dynamics, including the resulting $f(R)$ Lagrangian, in order to assess the viability of the model. We find that the standard viability conditions are satisfied, namely the positivity of the relevant derivatives of $f(R)$ and the positivity of the effective squared mass. We also compared the $f(R)$ of our model with the Starobinsky one, finding the same functional behavior in the curvature range of interest, up to an additive constant. From a broader inflationary perspective, our scenario operates in a regime where the scalar field stays close to the GR limit, $\phi \simeq 1$, rather than taking the larger values often encountered in the Starobinsky and Appleby-Battye-Starobinsky models during inflation. The reheating dynamics are likewise qualitatively different: whereas Starobinsky-type models reheat through post-inflationary oscillations of the scalaron around the minimum of the effective potential, transferring energy to matter fields, in our case radiation is generated already during inflation via gravitational particle creation, and the end of inflation connects smoothly to a $\Lambda$CDM radiation-dominated evolution.

Finally, although the present work is restricted to a background-level analysis and the adopted observables are not sensitive to cosmological perturbations, we nevertheless compute the theoretical predictions for the primordial spectral parameters $n_s$ and $r$. We find that the predicted value of the scalar spectral index $n_s$ is compatible with the most recent Planck+ACT+SPT constraints, provided that the perturbations generating anisotropies observed in the CMB cross the Hubble horizon approximately $50$--$54$ e-foldings before the end of inflation. For this range the tensor-to-scalar ratio satisfy the condition of $r<10^{-3}$.

Overall, our model successfully generates an early de Sitter phase driven by modified $f(R)$ dynamics: the non-minimally coupled scalar field is responsible for the slow-rolling phase throughout inflation, particle creation admits the reheating phase and the predicted primordial power spectrum observables are in agreement with the most recent CMB constraints. The considered model can be also viewed as an eternal inflation scenario and therefore it guarantees a sufficiently large e-folding value to address the basic Standard Model paradoxes.
This framework thus provides a self-consistent description linking inflationary physics to late-time cosmological observables.

\clearpage
\section*{CRedIT statement} 
\textbf{Elisa Fazzari}: Investigation; Conceptualization; Visualization; Writing – original draft; Writing – review \& editing;
\textbf{Chiara De Leo}: Investigation; Methodology; Software development; Visualization; Writing – original draft; Writing – review \& editing;
\textbf{Giovanni Montani}: Conceptualization; Writing – original draft; Writing – review \& editing.
\textbf{Matteo Martinelli}: Methodology; Software supervision; Writing – review \& editing;
\textbf{Alessandro Melchiorri}: Methodology; Writing – review \& editing;
\textbf{Guadalupe Cañas-Herrera}: Software development;  Writing – review \& editing.

\section*{Acknowledgments}
E.F. and G.M. would like to thank Tiziano Schiavone for the useful advices and discussion on the theoretical model development. 
Authors E.F. and A.M. are supported by "Theoretical Astroparticle Physics" (TAsP), iniziativa specifica INFN. The work of E.F., C.D.L and A.M. was partially supported by the research grant number 2022E2J4RK “PANTHEON: Perspectives in Astroparticle and Neutrino THEory with Old and New messengers” under the program PRIN 2022 funded by the Italian Ministero dell’Università e della Ricerca (MUR). M.M. acknowledges funding by the Agenzia Spaziale Italiana (\textsc{asi}) under agreement n. 2024-10-HH.0 and support from INFN/Euclid Sezione di Roma. G.C.H. acknowledges support through the European Space Agency research fellowship programme. 
C.D.L. and M.M. acknowledge financial support from Sapienza Università di Roma, provided through Progetti Medi 2021 (Grant No. RM12117A51D5269B).
\noindent This work made use of Melodie, a computing infrastructure funded by the same project, and PLEIADI, a computing infrastructure installed and managed by INAF.

\appendix

\section{Preliminary analysis on free parameters} \label{sec:appA}

\begin{figure}[H]
    \centering
    \includegraphics[width=0.75\linewidth]{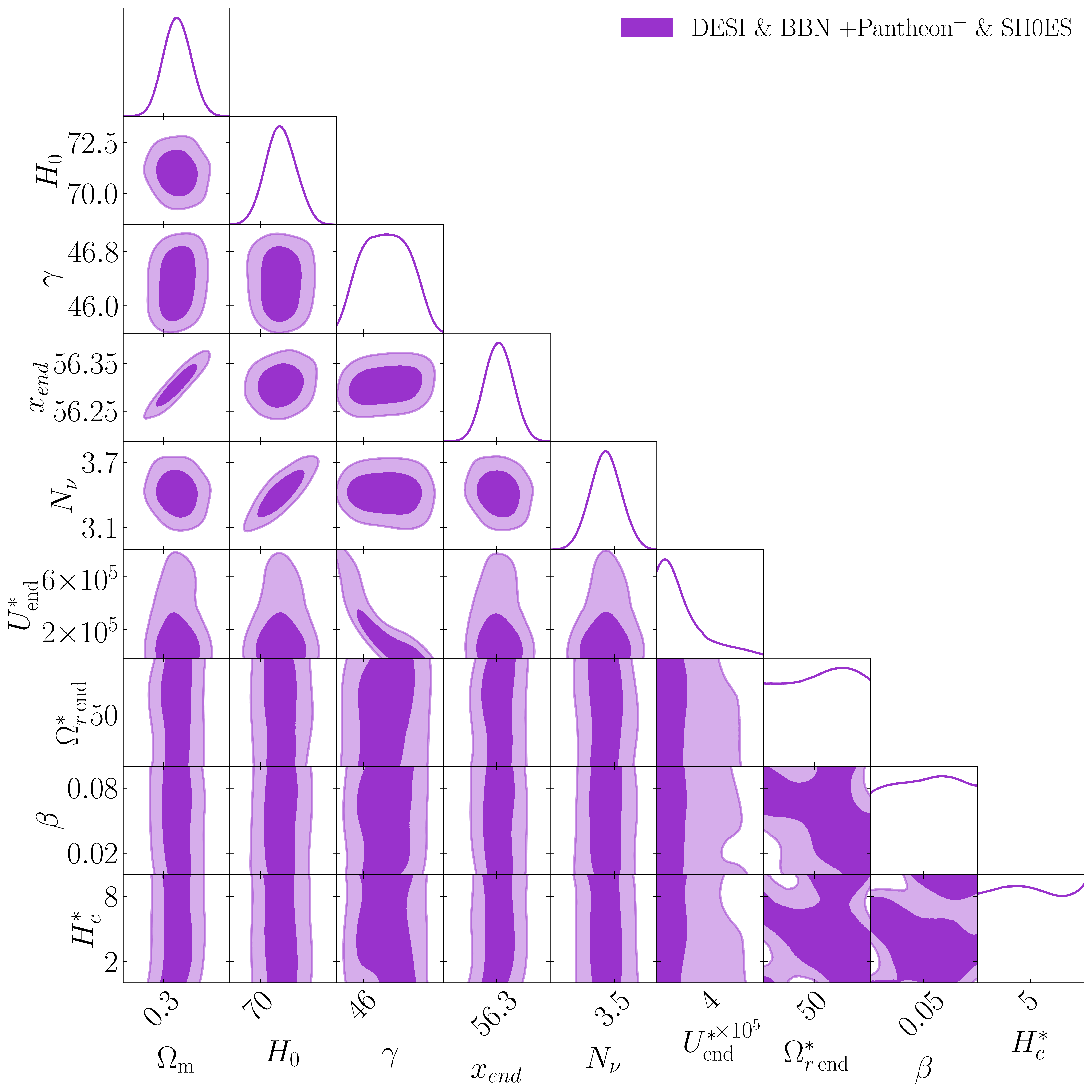}
    \caption{One and two dimensional marginalized constraints on the cosmological and model parameters considering also $\Omega^*_{r_\mathrm{end}}$, $\beta$ and $H_c^*$, using the full dataset DESI+BBN+Pantheon$^+$+SH0ES.}
    \label{fig:contour_flat}
\end{figure}

In this Appendix we present a preliminary theoretical analysis of the parameters that do not impact the background evolution and are therefore not constrained by the datasets considered in this work. Since our analysis is restricted to background observables, any parameter that does not modify the background dynamics remains unconstrained in the sampling procedure. This is explicitly illustrated in \autoref{fig:contour_flat}, where we show that $\beta$, $H_c^{*}$, and $\Omega^{*}_{r_{\mathrm{end}}}$, sampled with the flat prior ranges specified in \autoref{tab:prior}, exhibit completely flat posterior distributions. 
Moreover, their inclusion does not affect the constraints on the other cosmological parameters, confirming that the datasets used in this work are insensitive to them.
Motivated by these considerations, we fix these parameters to representative values supported by theoretical consistency.

\begin{figure}[H]
    \centering
    \begin{subfigure}[b]{0.7\linewidth}
        \centering
        \includegraphics[width=\linewidth]{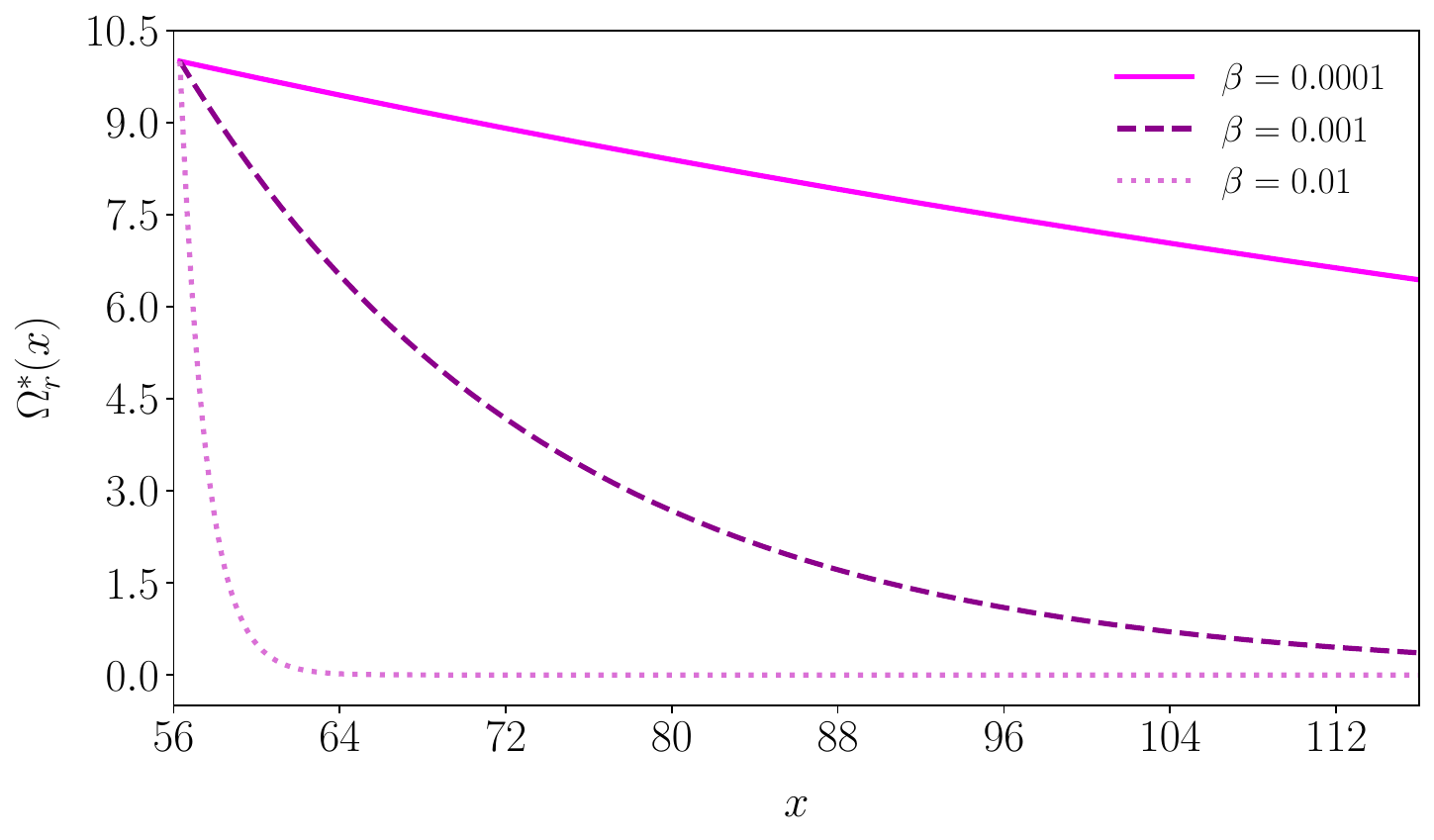}
        \caption{}
        \label{fig:Omegar_beta}
    \end{subfigure}
    \hfill
    \begin{subfigure}[b]{0.7\linewidth}
        \centering
        \includegraphics[width=\linewidth]{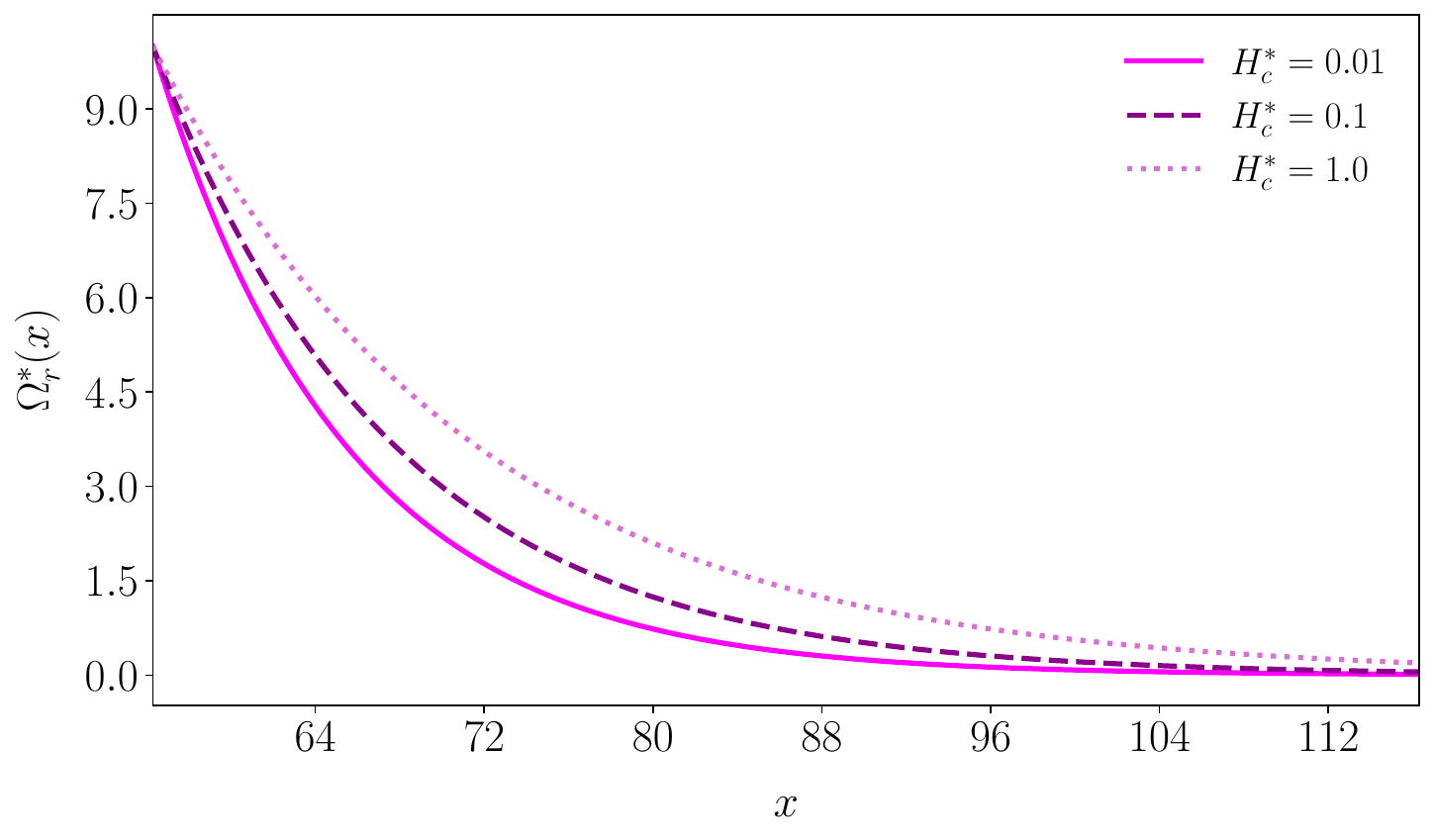}
        \caption{}
        \label{fig:Omegar_Hc}
    \end{subfigure}
    \caption{\textbf{(a)} Effect of $\beta$ parameter of particle creation on density radiation.  We show the case of $\beta=[10^{-4},10^{-3},10^{-2}]$ respectively in magenta (solid line), purple (dashed line), orchid (dash-dot line).
    \textbf{(b)} Impact of $H_c^{*}$ on $\Omega^*_r(x)$. We show the case of $H_c^{*}=[1.0,0.1, 0.01]$ respectively in magenta (solid line), purple (dashed line), orchid (dash-dot line).}
    \label{fig:Omegar_par}
\end{figure}

\begin{figure}[H]
    \centering
    \begin{subfigure}[b]{0.7\linewidth}
        \centering
        \includegraphics[width=\linewidth]{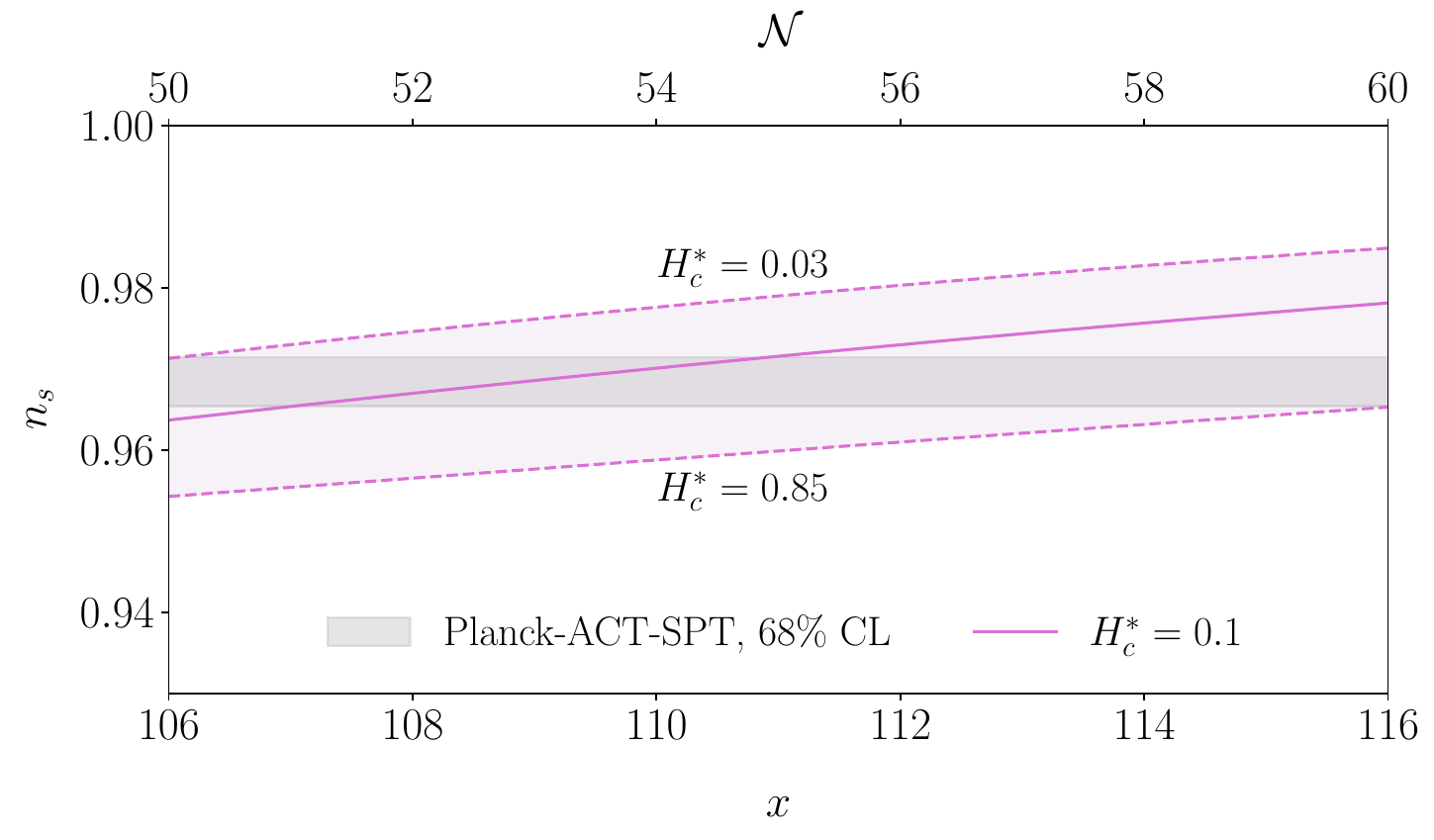}
        \caption{}
        \label{fig:ns_Hc}
    \end{subfigure}
    
    \begin{subfigure}[b]{0.7\linewidth}
        \centering
        \includegraphics[width=\linewidth]{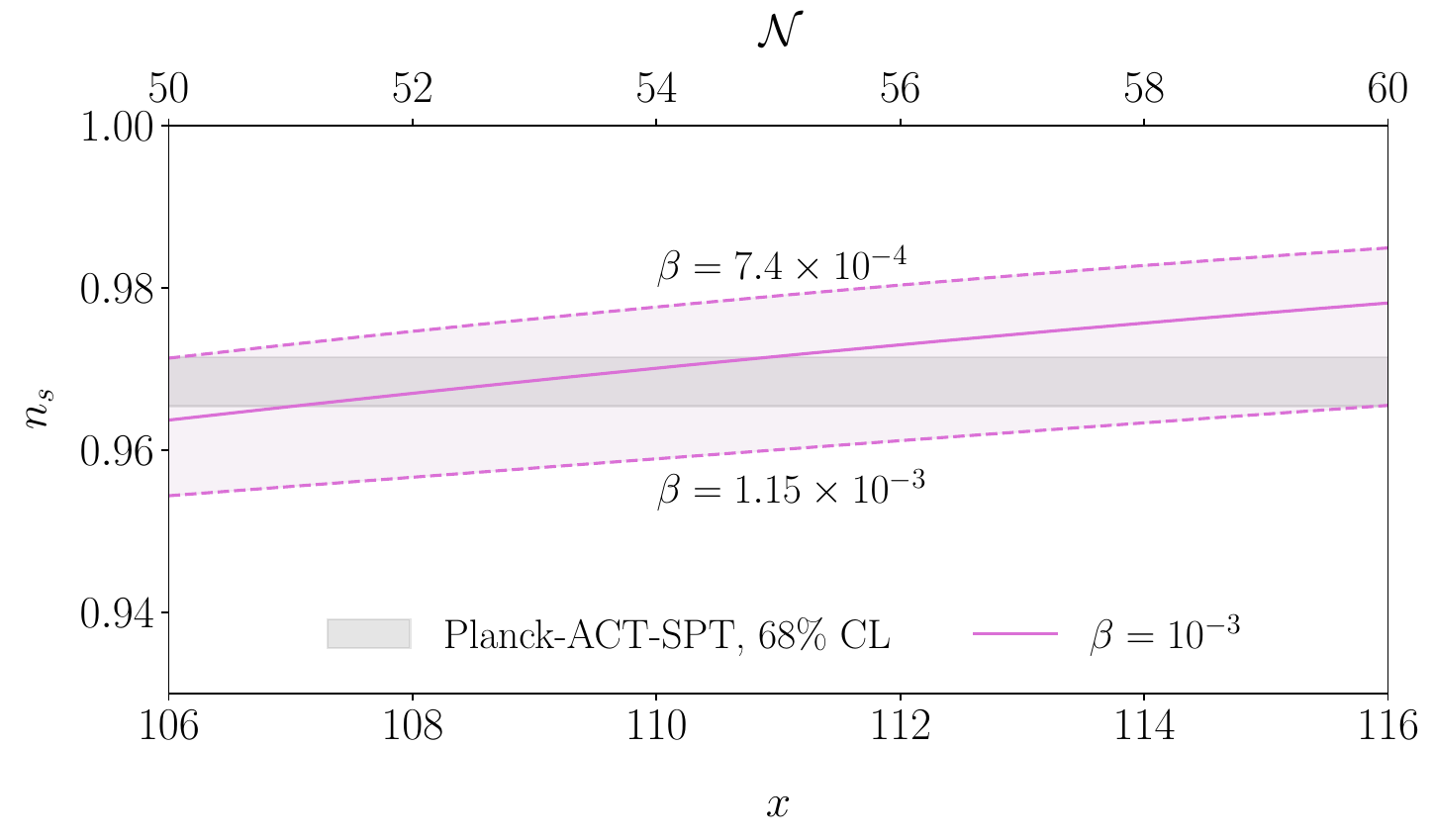}
        \caption{}
        \label{fig:ns_beta}
    \end{subfigure}
    
    \begin{subfigure}[b]{0.7\linewidth}
        \centering
        \includegraphics[width=\linewidth]{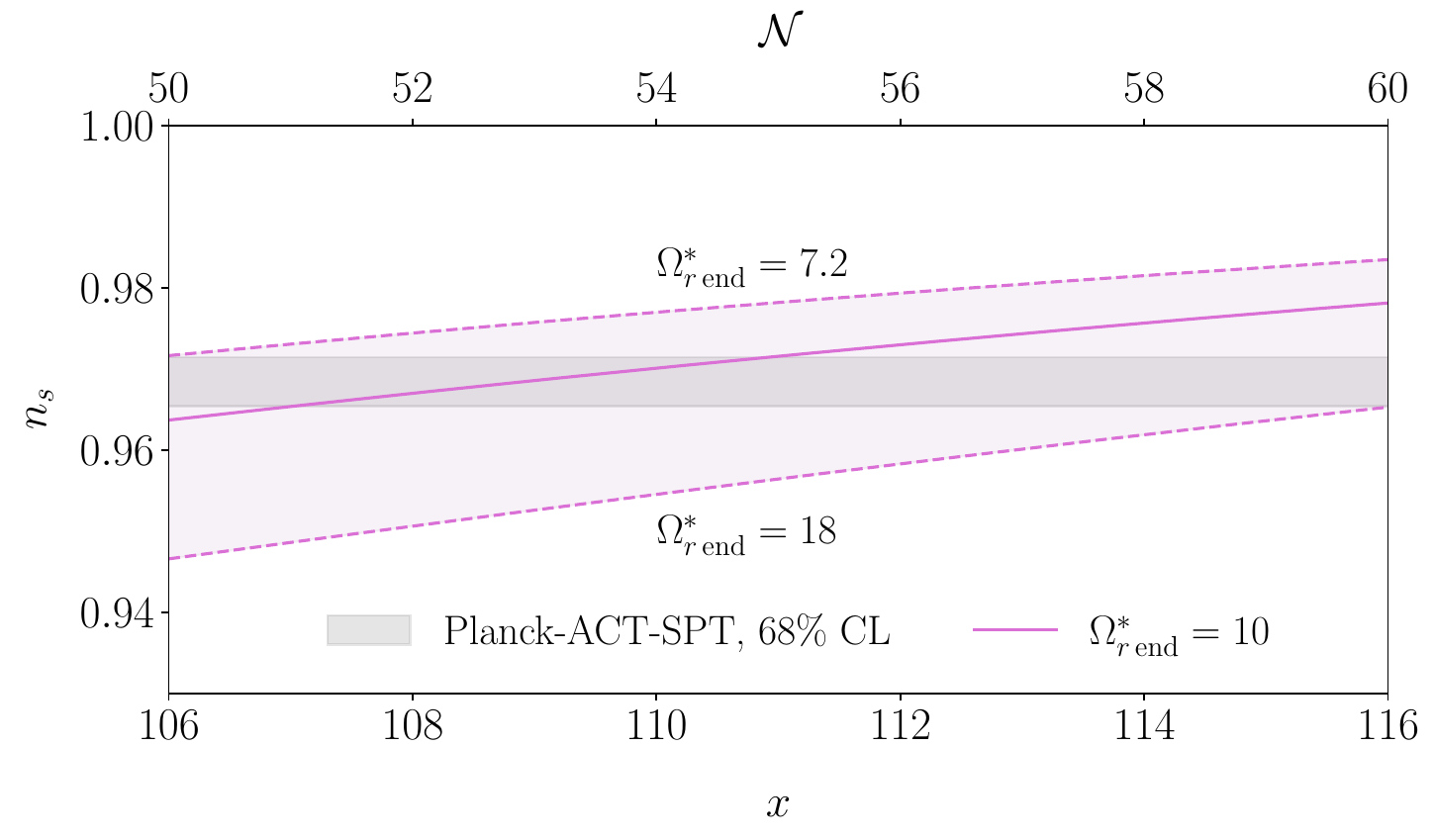}
        \caption{}
        \label{fig:ns_Omegar}
    \end{subfigure}
\caption{Sensitivity of the predicted scalar spectral index $n_s$ to variations in $H_c^*$ \textbf{(a)}, $\beta$ \textbf{(b)}, and $\Omega^*_{r_{\mathrm{end}}}$ \textbf{(c)}. In each panel, the allowed parameter range is determined by requiring consistency with the Planck+ACT+SPT $68\%$ C.L. constraint (gray band) \cite{SPT-3G:2025bzu}. When not varied, the parameters are fixed to $\beta=10^{-3}$, $\Omega^*_{r_{\mathrm{end}}}=10$, and $H_c^*=0.1$. The solid vertical line denotes the fiducial value adopted in the main analysis.}

    \label{fig:ns_par}
\end{figure}

\begin{figure}[H]
    \centering
    \includegraphics[width=0.75\linewidth]{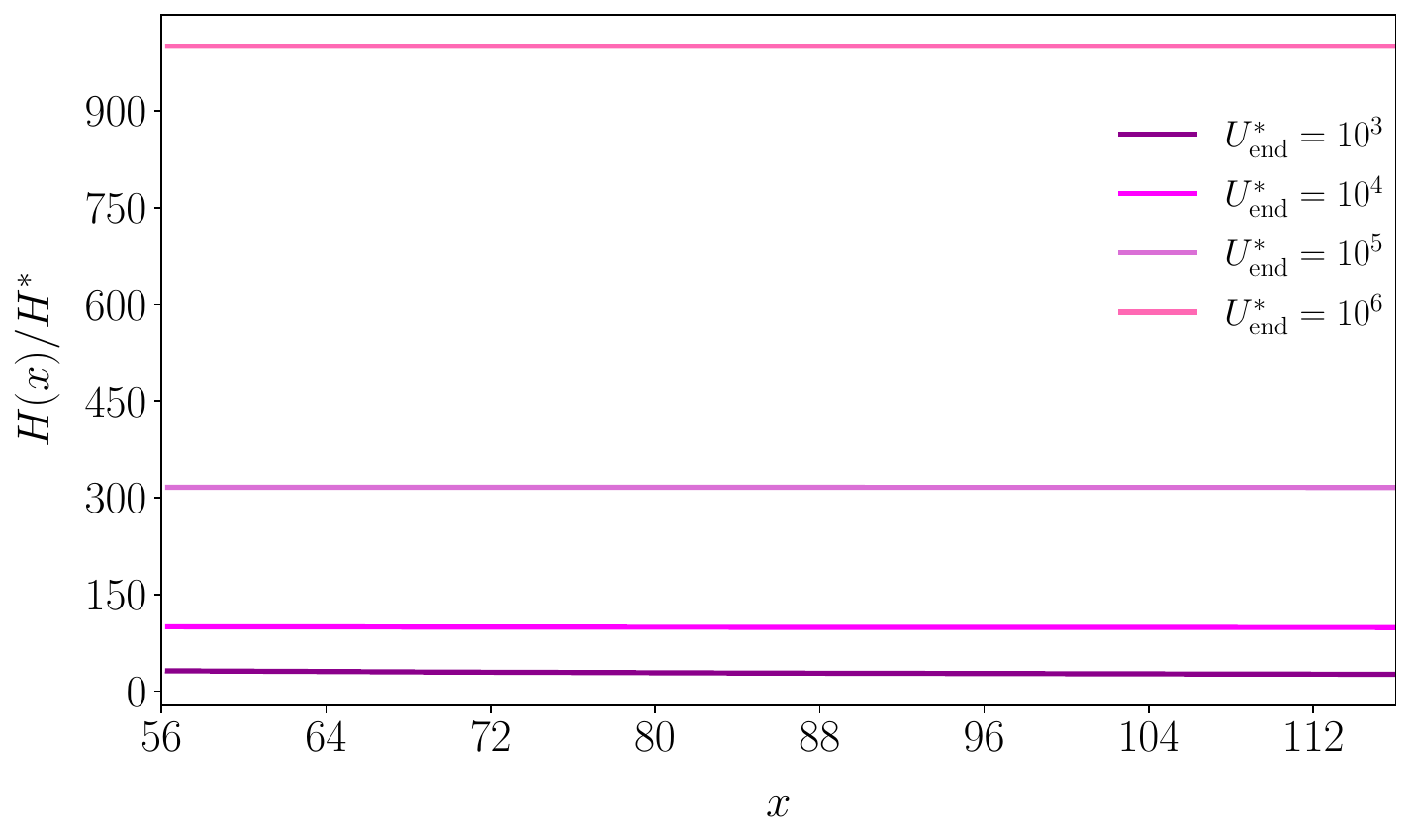}
    \caption{Impact of $U^*_{\rm end}$ on $H(x)$. We show the case of $U^*_{\rm end}=[10^3,10^4,10^5, 10^6]$.}
    \label{fig:H_Ustar}
\end{figure}

In the following, we first discuss the impact of $\beta$, $H_c^*$, and $\Omega^*_{r_{\mathrm{end}}}$ on the radiation particle production mechanism, and subsequently analyze their effect on the primordial power spectrum observables.
From this point onward, we adopt as fiducial values $x_{\mathrm{end}} = 56, \, H^* = 10^{46}, \, U^*_{\mathrm{end}} = 10^5, \, \phi_{\mathrm{end}} = 1.$
In \autoref{fig:Omegar_beta} we show the impact of the parameter $\beta$ on radiation particle production, exploring the range $\beta \in [10^{-4}, 10^{-2}]$. We find that $\beta$ controls the efficiency of the particle creation mechanism, although the considered values all lead to physically viable evolutions. In \autoref{fig:Omegar_Hc} we investigate the effect of varying $H_c^*$ within the range $H_c^* \in [0.01, 1]$. We observe that this parameter does not significantly alter the qualitative behavior of $\Omega^*_r(x)$, but mainly affects the overall trend of the radiation component. We do not explicitly show the impact of $\Omega^*_{r_{\mathrm{end}}}$, as its role is straightforward: being the boundary condition of Eq.~\eqref{eq:Omega_r_star}, it only determines the final value reached by $\Omega^*_r(x)$ without modifying its dynamical behavior.
Building on these results, we then examine how $H_c^*$, $\beta$, and $\Omega^*_{r_{\mathrm{end}}}$ influence the predicted scalar spectral index $n_s$. \autoref{fig:ns_par} displays, for each parameter, the range of values that keeps $n_s$ within the Planck+ACT+SPT $68\%$ C.L. constraint \cite{SPT-3G:2025bzu}, assuming the standard e-folding interval $\mathcal{N} \in (50,60)$. Based on this analysis, we adopt the fiducial values $
\beta = 10^{-3}, \, 
\Omega^*_{r_{\mathrm{end}}} = 10, \, 
H_c^* = 0.1,$ as this combination yields values of $n_s$ consistent with the Planck+ACT+SPT constraint across the relevant e-folding range.
For completeness, we also discuss the role of $U^*_{\rm end}$. We show that its variation primarily rescales the overall amplitude of the expansion rate without altering the qualitative background evolution. Its effect on the Hubble function is illustrated in \autoref{fig:H_Ustar}, confirming that it does not introduce additional structure in the background dynamics beyond a normalization shift.

\section{Physical viability of the model}\label{sec: app_rec}

\begin{figure}[h!]
    \centering
    \includegraphics[width=1.0\linewidth]{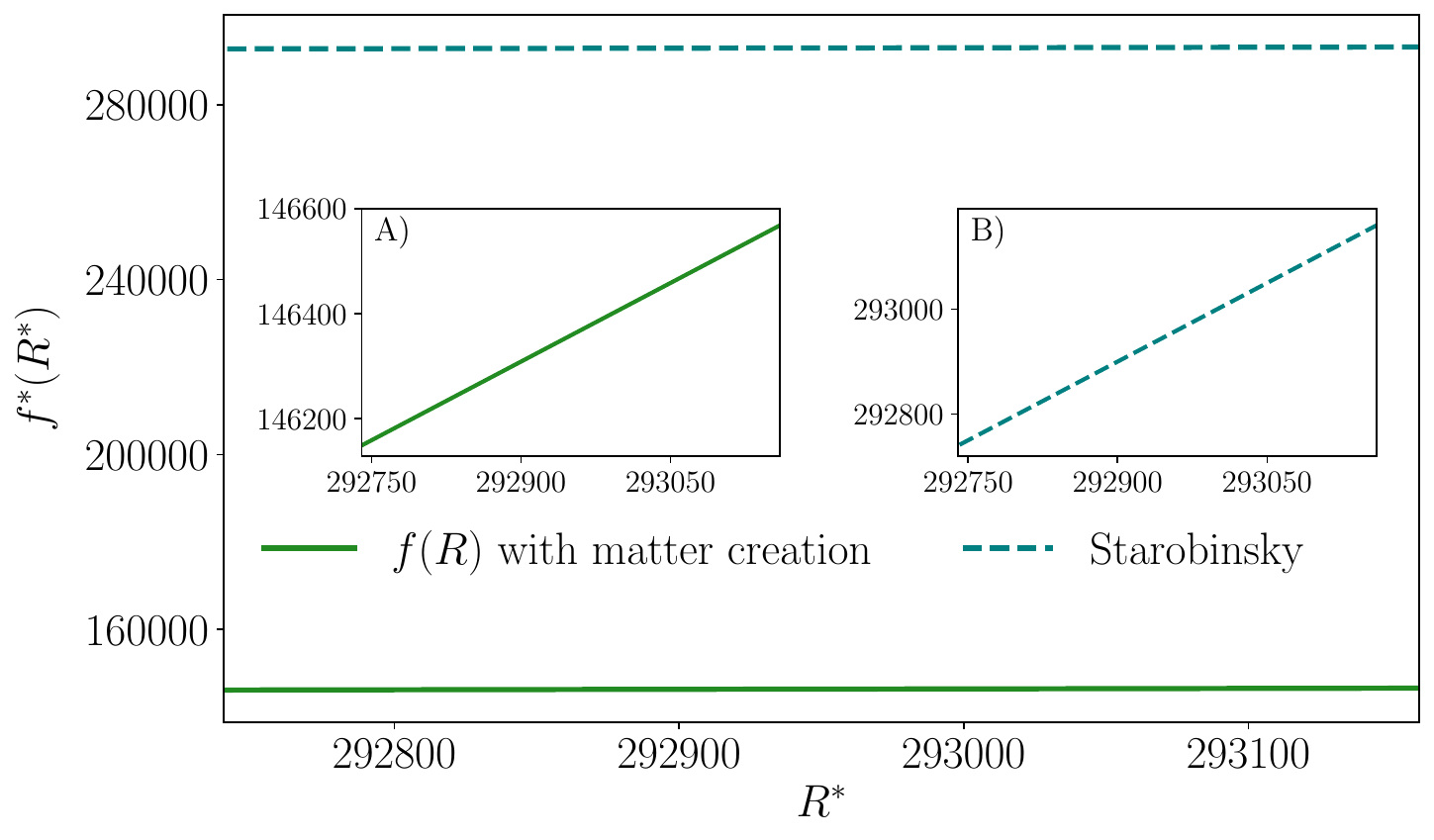}
    \caption{Comparison between our reconstructed normalized lagrangian denisty over the range of $R^*$ corresponding to the $x$ interval explored in the dynamics, obtained using the best-fit parameters in \autoref{tab:bestfit}, (solid line) and the Starobinsky model (dashed line).}
    \label{fig:fR_vs_Starobinski}
\end{figure}

\begin{figure}[h!]
    \centering
        \includegraphics[width=1.0\linewidth]{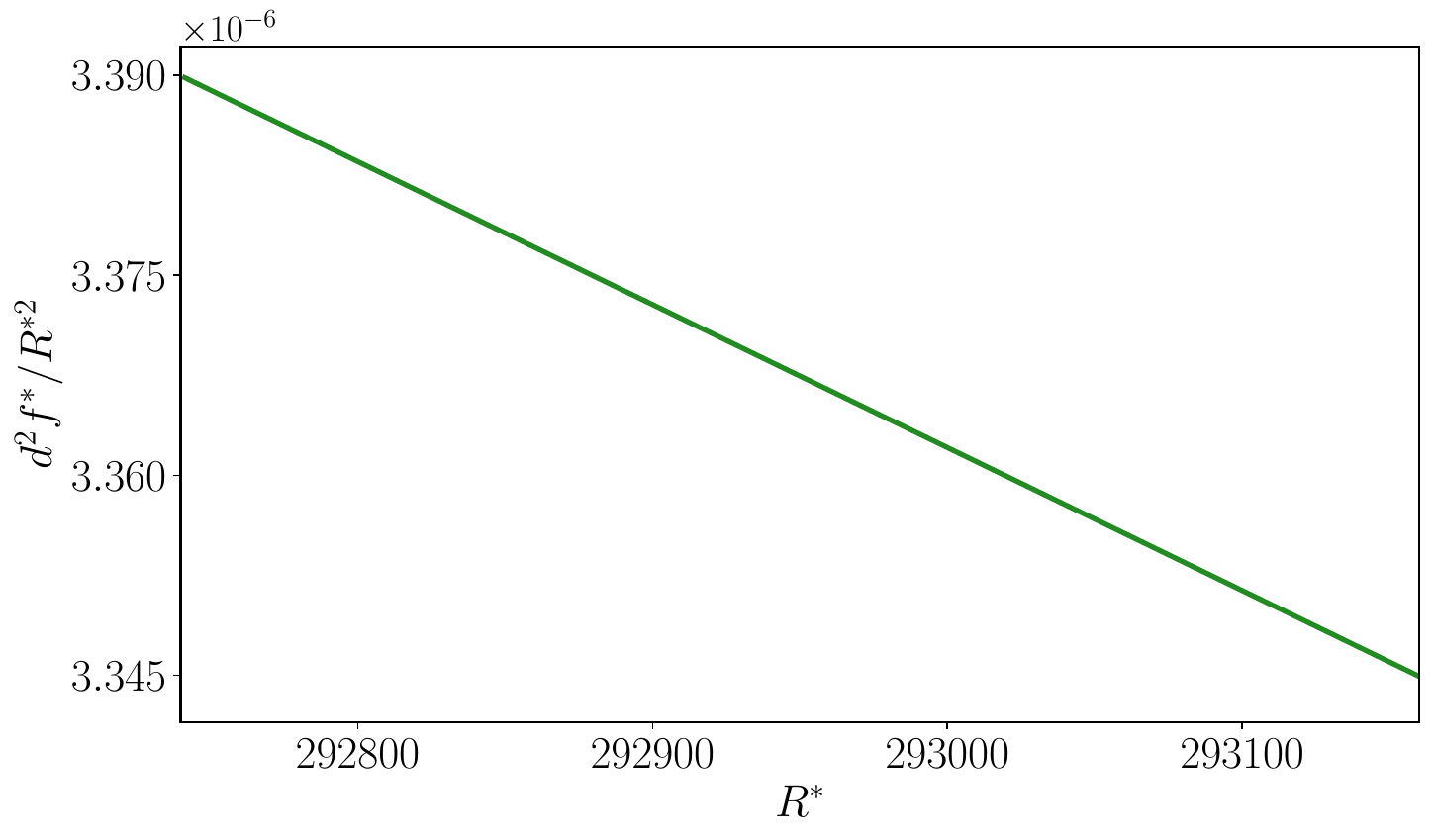}
    \caption{Second derivative of Langrangian density over the range of $R^*$ corresponding to the $x$ interval explored in the dynamics. This is obtained using the best-fit values in \autoref{tab:bestfit}.}
    \label{fig:f_RR}
\end{figure}

In this Appendix section, we assess the physical viability of the reconstructed $f(R)$ model by numerically deriving the corresponding Lagrangian density $f(R)$. The reconstruction is performed starting from the background solutions for $\phi(x)$ and $V^*(x)$, together with the normalized Ricci scalar $R^*(x) \equiv R/(6H^{*2})$. The latter is obtained from the relation $R(x) = 12H^{2}(x) - 3 H^{\prime 2}(x)$, which is equivalent to Eq.~\eqref{eq:vincolo_FLRW}.
We then invert these relations to express $\phi$ and $V^*$ as functions of $R^*$, and, using the definition in Eq.~\eqref{eq:fR_def} we obtain the normalized Lagrangian $f^*\equiv f/(6H^{*^2})$. Concerning physical viability, the reconstructed function $f^*(R^*)$, shown in inset A of \autoref{fig:fR_vs_Starobinski}, is found to be monotonically increasing, thereby satisfying the condition $f_R > 0$. Furthermore, as illustrated in \autoref{fig:f_RR}, its second derivative is positive, ensuring stability against the Dolgov–Kawasaki instability.

In \autoref{fig:fR_vs_Starobinski} we also compare our result with the Starobinsky lagrangian density \cite{Starobinsky} normalized consistently with the definitions adopted above, i.e. $f^*(R^*)=R^*+\alpha^* R{^*{^2}}$ where $\alpha^*=\left({H^*}/{M}\right)^2=10^{-17}$, using $M=10^{13}$ GeV \cite{DeFelice:2010aj} and $H^*=10^4$ GeV from our best-fit values. In the considered regime, our model and the Starobinsky one differ only by a constant offset. This is expected, since the potential adopted in our model contains a constant term, which is absent in the Starobinsky case.

\begin{figure}
    \centering
    \includegraphics[width=0.8\linewidth]{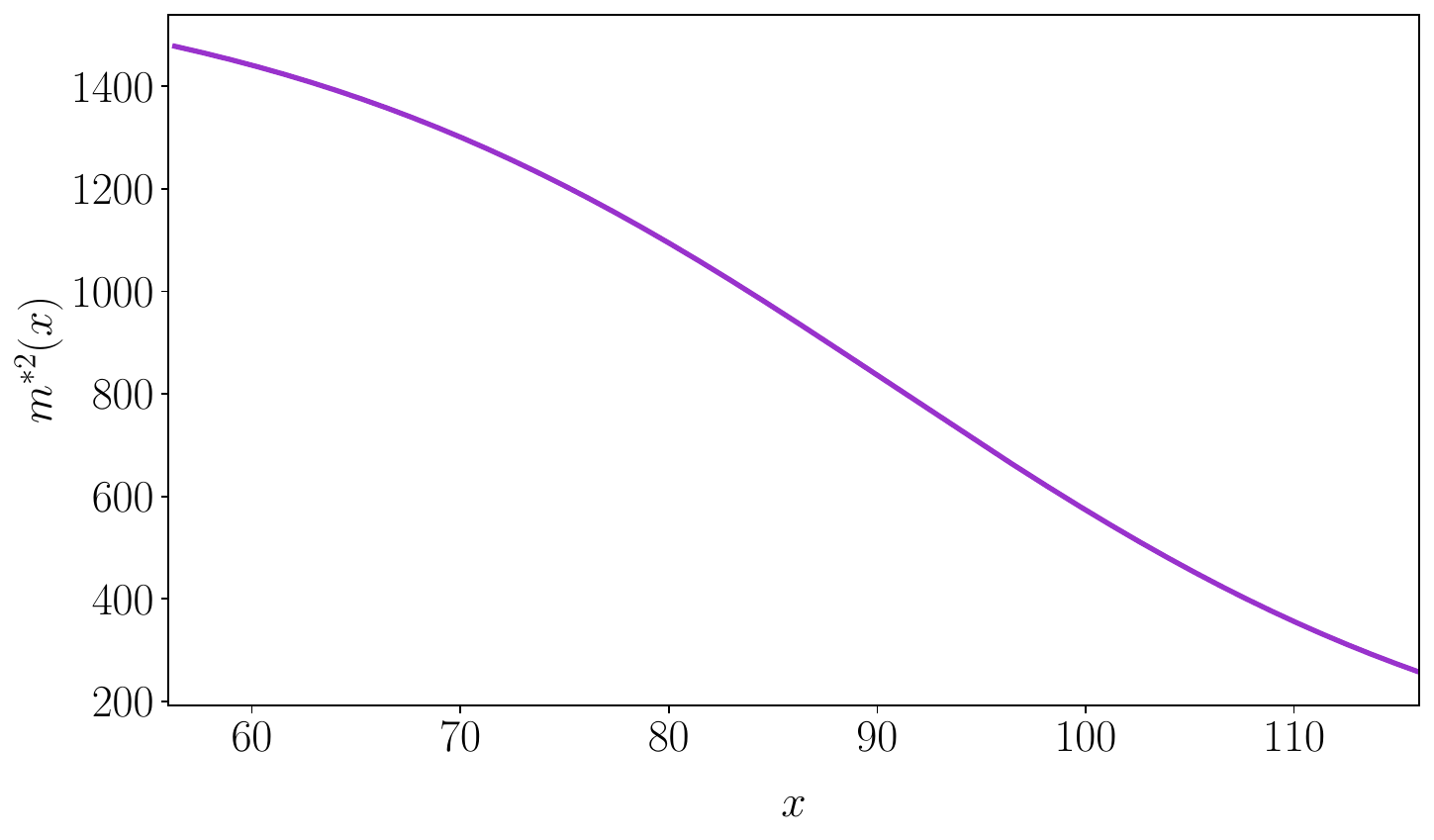}
    \caption{Squared mass of the scalar field $\phi$ computed using the best-fit values reported in \autoref{tab:bestfit}.}
    \label{fig:m2}
\end{figure}

We also evaluated the normalized squared mass of the scalar field using Eq.~\eqref{eq:m2_phi} rewritten in terms of the $x$ variable:
\begin{equation}
     m^{*^2}(x) \equiv \frac{m^2}{6H^{*^2}}= \frac{1}{3} \left(\phi(x) \frac{U^{*^{\prime\prime}}(x) \phi^{\prime}(x) - U^{*^\prime}(x) \phi^{\prime\prime}(x)}{\phi^{\prime}(x)^3} -  U^{*^\prime}(x) \phi^{\prime}(x)\right) \,.
     \label{eq:m2_X}
\end{equation}
\autoref{fig:m2} confirms that throughout the inflationary regime considered here the model does not develop a tachyonic instability, since the squared mass $m^{*^2}$ remains positive.

\newpage
\section{Non viable alternative ansatz} \label{sec: appB}
\begin{figure}[h!]
    \centering
    \begin{subfigure}[b]{0.48\linewidth}
        \centering
        \includegraphics[width=\linewidth]{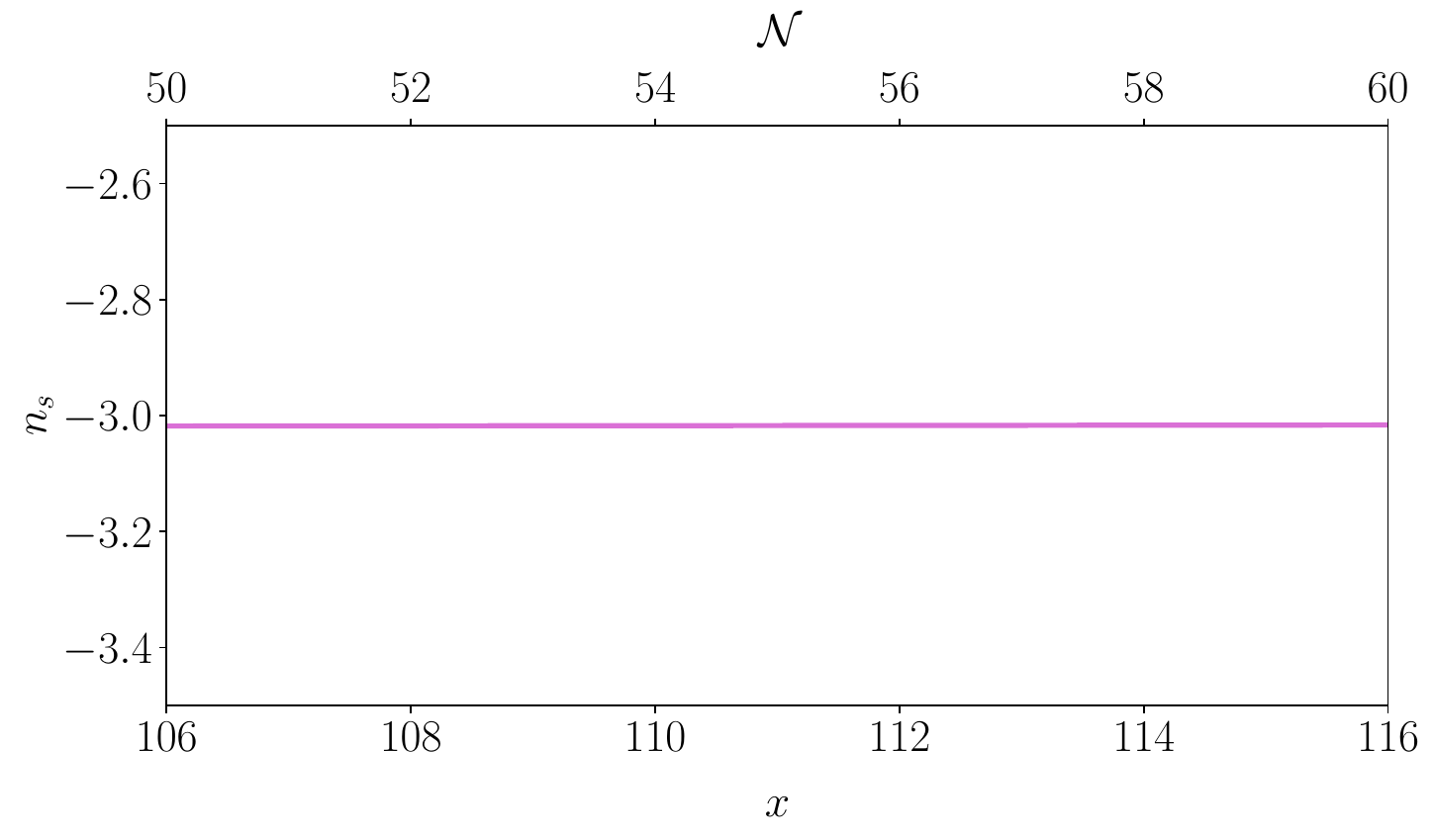}
        \caption{}
        \label{fig:ns_app}
    \end{subfigure}
    \hfill
    \begin{subfigure}[b]{0.48\linewidth}
        \centering
        \includegraphics[width=\linewidth]{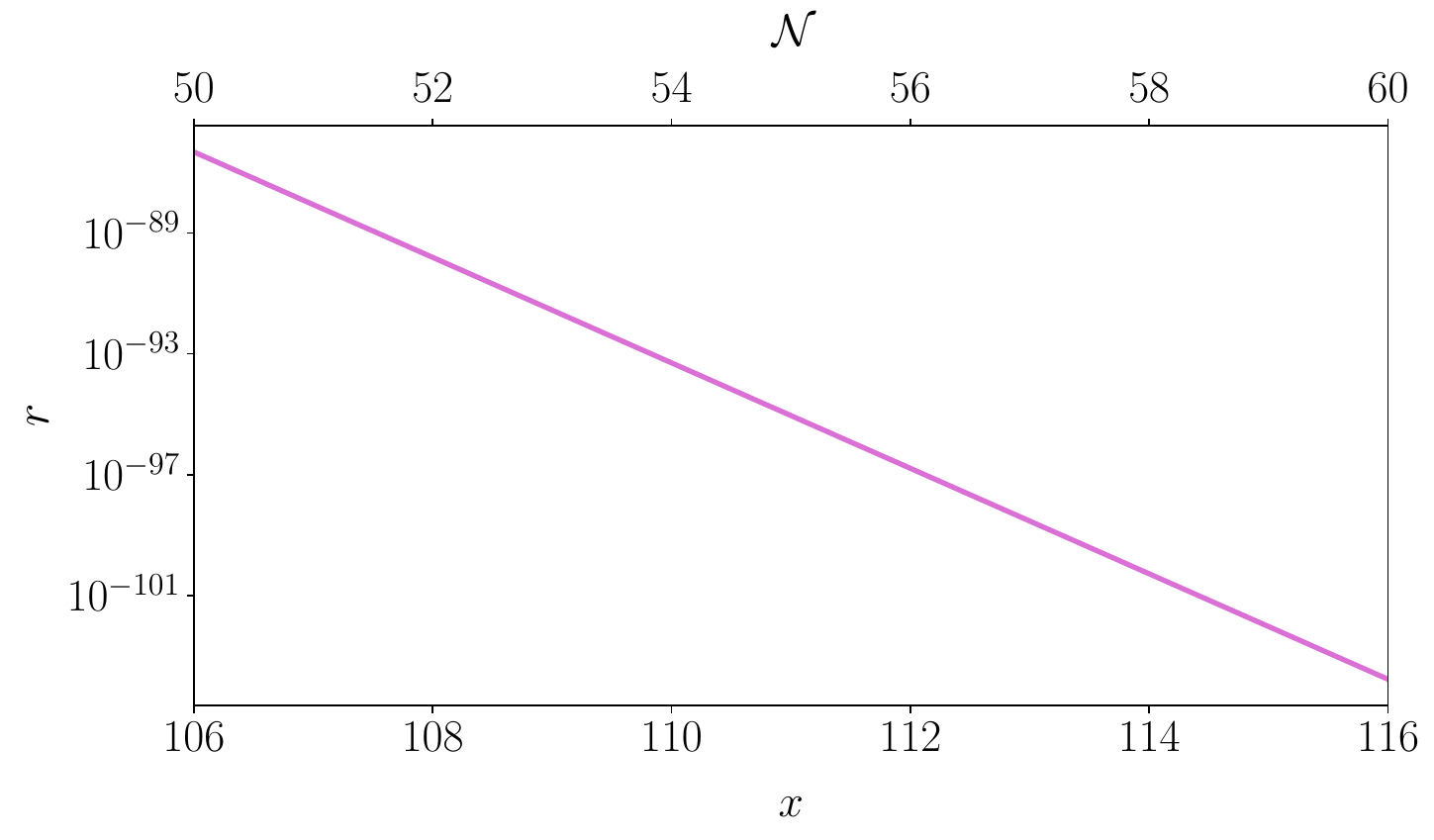}
        \caption{}
        \label{fig:r_app}
    \end{subfigure}
    \caption{Spectral index (a) and tensor-to-scalar ration (b) predicted using the best-fit values of DESI+BBN+Pantheon$^+$+SH0ES for the alternative ansatz to the cosmological dynamics.}
    \label{fig:app_spectral_obs}
\end{figure}

\begin{figure}
    \centering
    \includegraphics[width=1.0\linewidth]{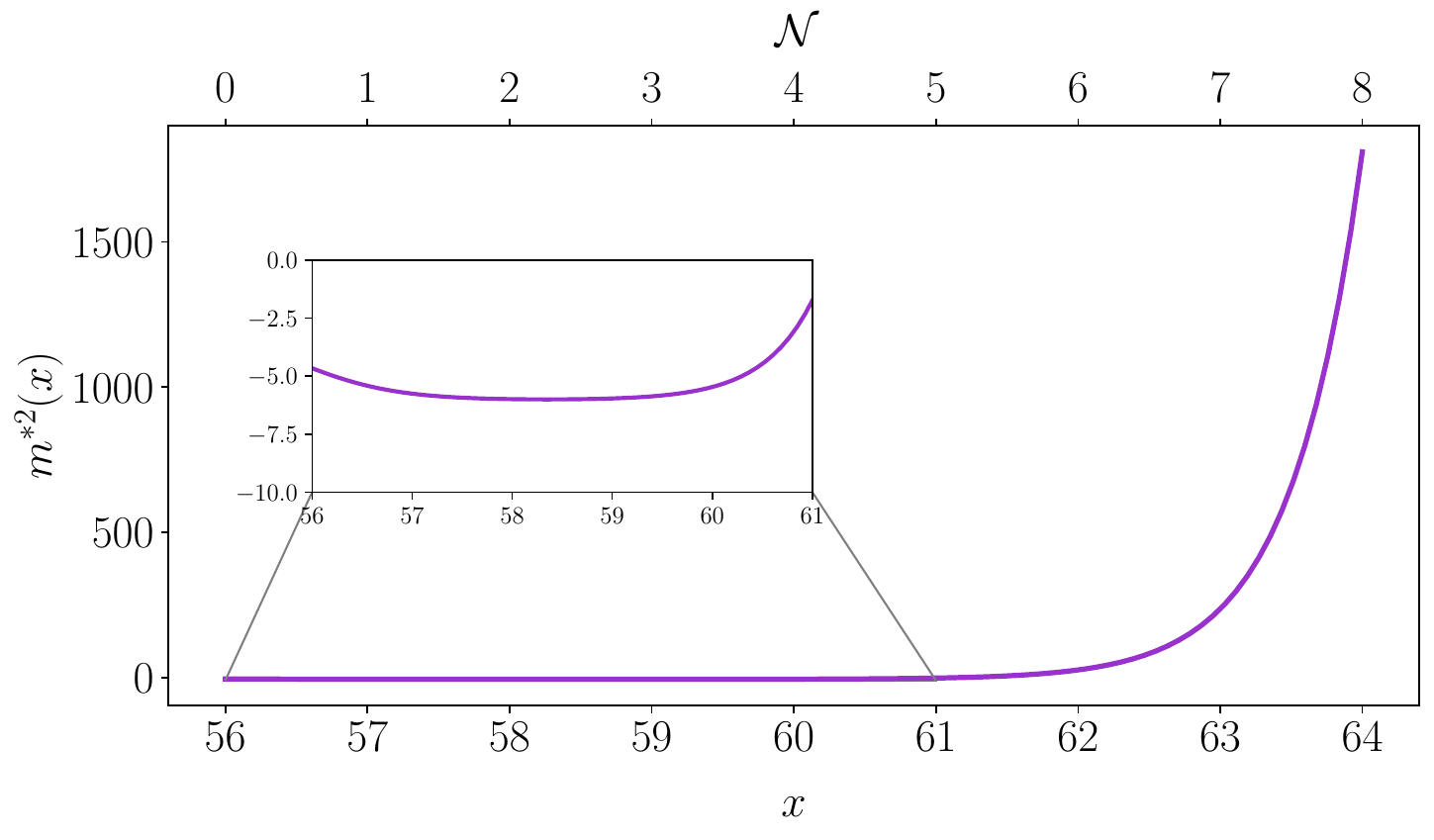}
    \caption{Squared mass for the scalar field negative for values of $x$ near $x_{end}$ using best-fit values of DESI+BBN+Pantheon$^+$+SH0ES for the alternative ansatz to the cosmological dynamics.}
    \label{fig:m2_app}
\end{figure}

Here, we discuss the other possible \textit{ansatz} to Eq. \eqref{eq:dinamica_2}, showing why this is not a viable model for the inflationary dynamics due to the fact that it predicts a spectral index above the observed range with the most recent data. \\
We can obtain the other solution decoupling the Friedmann equation \eqref{eq:dinamica_2} into the following two equations:
\begin{subequations}
\label{eq:app_disaccoppiamento}
\begin{equation}
H^2=\frac{\chi}{3\phi} \left(\rho_r+\rho_{\Lambda}\right) \,,
\label{eq:app_sm_1}
\end{equation}
\noindent \text{and}
\begin{equation}
U=-6H^2\frac{d\phi}{dx} \,.
\label{eq:app_sm_2}
\end{equation}
\end{subequations}
By dividing Eq.~\eqref{eq:vincolo_2} by Eq.~\eqref{eq:app_sm_2}, we obtain the equation:
\begin{equation}
    \frac{d \ln U}{dx}=-2+\frac{d \ln H}{dx} \,,
\end{equation}
which admits the solution:
\begin{equation}
    U=\frac{C H}{e^{2x}}\,,
    \label{eq:app_sm3}
\end{equation}
where $C$ is an integration constant that we assume to be positive. \\
Now, comparing the equation above with Eq.~\eqref{eq:app_sm_2} and taking into account the expression \eqref{eq:app_sm_1} we obtain the differential equation for the scalar field:
\begin{equation}
    \phi'=-\frac{C}{6e^{2x}} \sqrt{\frac{\phi}{\frac{\chi}{3}\left(\rho_r+\rho_{\Lambda}\right)}}.
    \label{eq:app_phi_prime}
\end{equation}
Similarly to what done in \autoref{sec: formulation}, we use the same normalization constant defining: 
\begin{equation}
    H^{*^2}=\frac{\chi}{3} \rho_{\Lambda} \;,
    \label{eq: app_H_star}
\end{equation}
Consequently, the equations governing the cosmological dynamics, namely Eqs.\eqref{eq:app_sm_1}, \eqref{eq:mod_continuity_rad}, \eqref{eq:app_phi_prime}, transform into:
\begin{subequations}
\begin{equation}
H^2 = \frac{H^{*^2}}{\phi} \left( \Omega_{r^*} + 1 \right) \,,
\label{eq:app_Hubble_mod}
\end{equation}
\begin{equation}
    \ \Omega_{r^*}' = 4 \Omega_{r^*} \left(1 - \frac{\left(\frac{1}{\phi} \left(  \Omega_{r^*} + 1 \right)\right)^{\beta}}{H_{c*}^{2\beta}} \right) \,,
\end{equation}
\begin{equation}
    \phi' = -\frac{C}{6e^{2x}H^{*}} \sqrt{\frac{\phi}{\Omega_{r^*} + 1 }} \,,
\end{equation}
\begin{equation}
    V^*=1 + U^* \,,
\end{equation}
\label{eq:app_sist_finale}
\end{subequations}
\noindent where we introduced the normalized parameters:
\begin{equation}
	\Omega_{r^*}\equiv \frac{\rho _r}{\rho_{\Lambda}}\, \, ,\, H_{c*} \equiv \frac{H_c}{H^*} \, , \, V^*(\phi) \equiv \frac{V(\phi)}{6H^{*^2}}
	\,.
	\label{eq:app_par_norm}
\end{equation}
\noindent These coupled differential equations for the variables $\phi(x)$ and $\Omega_{r^*}(x)$ can be solved numerically by imposing boundary conditions consistent with that one chosen in \autoref{sec: formulation}:
\begin{equation}
    \Omega_{r^*}(x_{\rm end})=10 \quad \mathrm{and} \quad \phi(x_{\rm end})=1 \,,
    \label{eq: app_boundary_cond}
\end{equation} 
where $x_{\rm end}$ represents the redshift at which the inflationary phase ends. 
For this case we repeat the same analysis presented in the main text, sampling over the same cosmological and model parameters. 
Taking as reference the best-fit obtained in the DESI $\&$ BBN + Pantheon$^+$ $\&$ SH0ES case, we find out that the fitted free parameters obtained with the same method predicts a scalar-index well above the observed range by Planck-ACT-SPT \cite{SPT-3G:2025bzu}, as shown in \autoref{fig:app_spectral_obs}. Furthermore, we compute the squared mass of the non minimally scalar field using Eq. \ref{eq:m2_X} and, and as show in \autoref{fig:m2_app}, we find that it is negative near the matching point with the $\Lambda$CDM model, indicating the presence of a tachyonic instability. For these reasons, we reject this model.

\newpage
\bibliographystyle{JHEP}
\bibliography{biblio.bib}

@article{Sato:1981qmu,
    author = "Sato, Katsuhiko",
    title = "{First-order phase transition of a vacuum and the expansion of the Universe}",
    doi = "10.1093/mnras/195.3.467",
    journal = "Mon. Not. Roy. Astron. Soc.",
    volume = "195",
    number = "3",
    pages = "467--479",
    year = "1981"
}

@article{Odintsov:2026fyc,
    author = "Odintsov, S. D. and Oikonomou, V. K. and Sharov, G. S.",
    title = "{Viable F(R) scenarios unifying inflation with realistic dynamical dark energy}",
    eprint = "2601.06949",
    archivePrefix = "arXiv",
    primaryClass = "gr-qc",
    doi = "10.1016/j.jheap.2026.100579",
    journal = "JHEAp",
    volume = "52",
    pages = "100579",
    year = "2026"
}

@article{Oikonomou:2025qub,
    author = "Oikonomou, V. K.",
    title = "{Model Agnostic $F(R)$ Gravity Inflation}",
    eprint = "2504.00915",
    archivePrefix = "arXiv",
    primaryClass = "gr-qc",
    month = "4",
    year = "2025"
}

@article{Odintsov:2020thl,
    author = "Odintsov, S. D. and Oikonomou, V. K.",
    title = "{Inflationary attractors in F(R) gravity}",
    eprint = "2005.12804",
    archivePrefix = "arXiv",
    primaryClass = "gr-qc",
    doi = "10.1016/j.physletb.2020.135576",
    journal = "Phys. Lett. B",
    volume = "807",
    pages = "135576",
    year = "2020"
}

@article{Odintsov:2023weg,
    author = "Odintsov, Sergei D. and Oikonomou, Vasilis K. and Giannakoudi, Ifigeneia and Fronimos, Fotis P. and Lymperiadou, Eirini C.",
    title = "{Recent Advances in Inflation}",
    eprint = "2307.16308",
    archivePrefix = "arXiv",
    primaryClass = "gr-qc",
    doi = "10.3390/sym15091701",
    journal = "Symmetry",
    volume = "15",
    number = "9",
    pages = "1701",
    year = "2023"
}

@article{Oikonomou:2020oex,
    author = "Oikonomou, V. K.",
    title = "{Rescaled Einstein-Hilbert Gravity from $f(R)$ Gravity: Inflation, Dark Energy and the Swampland Criteria}",
    eprint = "2012.01312",
    archivePrefix = "arXiv",
    primaryClass = "gr-qc",
    doi = "10.1103/PhysRevD.103.124028",
    journal = "Phys. Rev. D",
    volume = "103",
    number = "12",
    pages = "124028",
    year = "2021"
}

@article{Kouniatalis:2025orn,
    author = "Kouniatalis, Gerasimos and Saridakis, Emmanuel N.",
    title = "{Inflation from a generalized exponential plateau: towards extra suppressed tensor-to-scalar ratios}",
    eprint = "2507.17721",
    archivePrefix = "arXiv",
    primaryClass = "astro-ph.CO",
    month = "7",
    year = "2025"
}

@article{Lopez:2025gfu,
    author = "L{\'o}pez, Samuel S{\'a}nchez and Terente D{\'\i}az, Jos{\'e} Jaime",
    title = "{Scalar-Induced Gravitational Waves in Palatini $f(R)$ Gravity}",
    eprint = "2505.13420",
    archivePrefix = "arXiv",
    primaryClass = "astro-ph.CO",
    month = "5",
    year = "2025"
}

@article{Kuralkar:2025zxr,
    author = "Kuralkar, Hardik Jitendra and Panda, Sukanta and Vidyarthi, Archit",
    title = "{Effective Starobinsky pre-inflation}",
    eprint = "2504.15061",
    archivePrefix = "arXiv",
    primaryClass = "gr-qc",
    month = "4",
    year = "2025"
}

@article{Gomes:2016cwj,
    author = "Gomes, Cl{\'a}udio and Rosa, Jo{\~a}o G. and Bertolami, Orfeu",
    title = "{Inflation in non-minimal matter-curvature coupling theories}",
    eprint = "1611.02124",
    archivePrefix = "arXiv",
    primaryClass = "gr-qc",
    doi = "10.1088/1475-7516/2017/06/021",
    journal = "JCAP",
    volume = "06",
    pages = "021",
    year = "2017"
}

@article{Gomes:2018uhv,
    author = "Gomes, Cl{\'a}udio and Bertolami, Orfeu and Rosa, Jo{\~a}o G.",
    title = "{Inflation with $Planck$ data: A survey of some exotic inflationary models}",
    eprint = "1803.08084",
    archivePrefix = "arXiv",
    primaryClass = "hep-th",
    doi = "10.1103/PhysRevD.97.104061",
    journal = "Phys. Rev. D",
    volume = "97",
    number = "10",
    pages = "104061",
    year = "2018"
}

@misc{whitepaper_cosmoverse,
      title="{The CosmoVerse White Paper: Addressing observational tensions in cosmology with systematics and fundamental physics}", 
      author={Di Valentino, Eleonora and Jackson Levi Said and Adam Riess and Agnieszka Pollo and Vivian Poulin and  Gómez-Valente, Adria and et al.},
      year={2025},
      eprint={2504.01669},
      archivePrefix={arXiv},
      primaryClass={astro-ph.CO},
      url={https://arxiv.org/abs/2504.01669}, 
}

@article{boomerang,
  title="{Cosmological parameters from the first results of Boomerang}",
  author={Lange, Andrew E and Ade, Peter AR and Bock, JJ and Bond, JR and Borrill, J and Boscaleri, A and Coble, K and Crill, BP and de Bernardis, Paolo and Farese, P and others},
  journal={Physical Review D},
  volume={63},
  number={4},
  pages={042001},
  year={2001},
  publisher={APS}
}

@article{Planck18,
   title="{Planck2018 results: VI. Cosmological parameters}",
   volume={641},
   ISSN={1432-0746},
   url={http://dx.doi.org/10.1051/0004-6361/201833910},
   DOI={10.1051/0004-6361/201833910},
   journal={Astronomy $\&$ Astrophysics},
   publisher={EDP Sciences},
   collaboration = "Planck",
   year={2020},
   month=sep, 
   pages={A6} 
}

@article{divalentino-review,
   title="{In the realm of the Hubble tension—a review of solutions}",
   volume={38},
   ISSN={1361-6382},
   url={http://dx.doi.org/10.1088/1361-6382/ac086d},
   DOI={10.1088/1361-6382/ac086d},
   number={15},
   journal={Classical and Quantum Gravity},
   publisher={IOP Publishing},
   author={Di Valentino, Eleonora and Mena, Olga and Pan, Supriya and Visinelli, Luca and Yang, Weiqiang and Melchiorri, Alessandro and Mota, David F and Riess, Adam G and Silk, Joseph},
   year={2021},
   month=jul, pages={153001} }

@article{efstathiou_planck,
  title="{The evidence for a spatially flat Universe}",
  author={Efstathiou, George and Gratton, Steven},
  journal={Monthly Notices of the Royal Astronomical Society: Letters},
  DOI={10.1093/mnrasl/slaa093},
  volume={496},
  number={1},
  pages={L91--L95},
  year={2020},
  publisher={Oxford University Press}
}

@article{divalentino_planck,
  title="{Planck evidence for a closed Universe and a possible crisis for cosmology}",
  author={Di Valentino, Eleonora and Melchiorri, Alessandro and Silk, Joseph},
  journal={Nature Astronomy},
  DOI={10.1038/s41550-019-0906-9},
  volume={4},
  number={2},
  pages={196--203},
  year={2020},
  publisher={Nature Publishing Group}
}

@article{fazzari_H0z,
    author = "Fazzari, E. and Dainotti, M. G. and Montani, G. and Melchiorri, A.",
    title = "{The effective running Hubble constant in SNe Ia as a marker for the dark energy nature}",
    eprint = "2506.04162",
    archivePrefix = "arXiv",
    primaryClass = "astro-ph.CO",
    month = "6",
    year = "2025"
}

@article{Nojiri-odintsov,
    author={Nojiri, Shin’ichi and Odintsov, Sergei D.},
    title = "{Unified cosmic history in modified gravity: From f(R) theory to Lorentz non-invariant models}",
    doi = "10.1016/j.physrep.2011.04.001",
    journal = "Physics Reports",
   publisher={Elsevier BV},
    volume = "505",
    pages = "59-144",
    year = "2011"
}

@article{Capozzielo-Delaurentis,
	doi = {10.1016/j.physrep.2011.09.003},
  
	url = {https://doi.org/10.1016%2Fj.physrep.2011.09.003},
  
	year = {2011},
	month = {dec},
  
	publisher = {Elsevier {BV}},
  
	volume = {509},
  
	number = {4-5},
  
	pages = {167--321},
  
	author = {Salvatore Capozziello and Mariafelicia De Laurentis},
  
	title = "{Extended Theories of Gravity}",
  
	journal = {Physics Reports}
}

@article{matcre_montanimary,
  title="{Modified gravity in the presence of matter creation: Scenario for the late Universe}",
  author={Montani, Giovanni and Carlevaro, Nakia and De Angelis, Mariaveronica},
  journal={Entropy},
  volume={26},
  number={8},
  pages={662},
  year={2024},
  publisher={MDPI}
}

@article{MONTANI_mary,
    author = "Montani, Giovanni and De Angelis, Mariaveronica and Dainotti, Maria Giovanna",
    title = "{Decay of dark energy into dark matter in a metric f(R) gravity: Effective running Hubble constant}",
    eprint = "2506.13288",
    archivePrefix = "arXiv",
    primaryClass = "astro-ph.CO",
    doi = "10.1016/j.dark.2025.101969",
    journal = "Phys. Dark Univ.",
    volume = "49",
    pages = "101969",
    year = "2025"
}

@misc{montani_hubbletension_1,
      title="{Exploring the Hubble tension with a late time Modified Gravity scenario}", 
      author={Luis A. Escamilla and Donatella Fiorucci and Giovanni Montani and Eleonora Di Valentino},
      year={2024},
      eprint={2408.04354},
      archivePrefix={arXiv},
      primaryClass={astro-ph.CO} 
}

@article{montani_hubbletension_2,
    author = "Schiavone, Tiziano and Montani, Giovanni",
    title = "{Evolution of an effective Hubble constant in f(R) modified gravity}",
    eprint = "2408.01410",
    archivePrefix = "arXiv",
    primaryClass = "gr-qc",
    doi = "10.1393/ncc/i2025-25105-3",
    journal = "Nuovo Cim. C",
    volume = "48",
    number = "3",
    pages = "105",
    year = "2025"
}

@article{CANDI,
doi = {10.1088/1475-7516/2025/11/001},
url = {https://doi.org/10.1088/1475-7516/2025/11/001},
year = {2025},
month = {nov},
publisher = {IOP Publishing},
volume = {2025},
number = {11},
pages = {001},
author = {De Leo, Chiara and Martinelli, Matteo and D'Agostino, Rocco and Gianfagna, Giulia and Martins, C.J.A.P.},
title = {Distinguishing distance duality breaking models using electromagnetic and gravitational waves measurements},
journal = {Journal of Cosmology and Astroparticle Physics}
}

@article{montani_hubbletension_4,
    author = "Montani, Giovanni and Carlevaro, Nakia and Escamilla, Luis A. and Di Valentino, Eleonora",
    title = "{Kinetic model for dark energy{\textemdash}dark matter interaction: Scenario for the hubble tension}",
    eprint = "2404.15977",
    archivePrefix = "arXiv",
    primaryClass = "gr-qc",
    doi = "10.1016/j.dark.2025.101848",
    journal = "Phys. Dark Univ.",
    volume = "48",
    pages = "101848",
    year = "2025"
}

@article{montani_hubbletension_6,
   title="{Slow-rolling scalar dynamics as solution for the Hubble tension}",
   volume={44},
   DOI={10.1016/j.dark.2024.101486},
   journal={Physics of the Dark Universe},
   publisher={Elsevier BV},
   author={Montani, Giovanni and Carlevaro, Nakia and Dainotti, Maria Giovanna},
   year={2024},
   month=may, pages={101486} 
}

@article{montani_hubbletension_8,
    author = "Schiavone, Tiziano and Montani, Giovanni and Bombacigno, Flavio",
    title = "{f(R) gravity in the Jordan frame as a paradigm for the Hubble tension}",
    eprint = "2211.16737",
    archivePrefix = "arXiv",
    primaryClass = "gr-qc",
    doi = "10.1093/mnrasl/slad041",
    journal = "Mon. Not. Roy. Astron. Soc.",
    volume = "522",
    number = "1",
    pages = "L72--L77",
    year = "2023"
}

@article{nautilus,
    author = {Lange, Johannes U},
    title = "{nautilus: boosting Bayesian importance nested sampling with deep learning}",
    journal = {Monthly Notices of the Royal Astronomical Society},
    volume = {525},
    number = {2},
    pages = {3181-3194},
    year = {2023},
    month = {08},
    doi = {10.1093/mnras/stad2441},
    url = {https://doi.org/10.1093/mnras/stad2441},
    eprint = {https://academic.oup.com/mnras/article-pdf/525/2/3181/51331635/stad2441.pdf},
}

@article{riotto2017lectures,
      title="{Inflation and the Theory of Cosmological Perturbations}", 
      author={Antonio Riotto},
      year={2017},
      eprint={hep-ph/0210162},
      archivePrefix={arXiv},
      primaryClass={hep-ph},
      url={https://arxiv.org/abs/hep-ph/0210162}, 
}

@book{weinberg,
    publisher = {Oxford University Press},
    author = "Weinberg, Steven",
    title = "{Cosmology}",
    isbn = "978-0-19-852682-7",
    year = "2008"
}

@book{KolbTurner,
    author = "Kolb, Edward W. and Turner, Michael S.",
    title = "{The Early Universe}",
    reportNumber = "FERMILAB-BOOK-1990-01",
    doi = "10.1201/9780429492860",
    isbn = "978-0-429-49286-0, 978-0-201-62674-2",
    publisher = "Taylor and Francis",
    volume = "69",
    month = "5",
    year = "2019"
}

@book{peacock_cosmology,
place={Cambridge}, 
title="{Cosmological Physics}",
publisher={Cambridge University Press},
author={Peacock, J. A.}, 
year={1998}
}

@article{Guth,
    author = "Guth, Alan H.",
    editor = "Fang, Li-Zhi and Ruffini, R.",
    title = "{The Inflationary Universe: A Possible Solution to the Horizon and Flatness Problems}",
    reportNumber = "SLAC-PUB-2576",
    doi = "10.1103/PhysRevD.23.347",
    journal = "Phys. Rev. D",
    volume = "23",
    pages = "347--356",
    year = "1981"
}

@article{LINDE,
title = "{Chaotic inflation}",
journal = {Physics Letters B},
volume = {129},
number = {3},
pages = {177-181},
year = {1983},
issn = {0370-2693},
doi = {https://doi.org/10.1016/0370-2693(83)90837-7},
url = {https://www.sciencedirect.com/science/article/pii/0370269383908377},
author = {A.D. Linde}
}

@article{Nearly_Starobinsky,
	doi = {10.1103/physrevd.89.023518},
  
	url = {https://doi.org/10.1103%2Fphysrevd.89.023518},
  
	year = {2014},
	month = {jan},
  
	publisher = {American Physical Society ({APS})},
  
	volume = {89},
  
	number = {2},
  
	author = {L. Sebastiani and G. Cognola and R. Myrzakulov and S.{\hspace{0.167em}
}D. Odintsov and S. Zerbini},
  
	title = "{Nearly Starobinsky inflation from modified gravity}",
  
	journal = {Physical Review D}
}

@article{giare_inflation,
    author = "Giar{\`e}, William",
    title = "{Inflation, the Hubble tension, and early dark energy: An alternative overview}",
    eprint = "2404.12779",
    archivePrefix = "arXiv",
    primaryClass = "astro-ph.CO",
    doi = "10.1103/PhysRevD.109.123545",
    journal = "Phys. Rev. D",
    volume = "109",
    number = "12",
    pages = "123545",
    year = "2024"
}

@article{Starobinsky,
    author = "Starobinsky, Alexei A.",
    editor = "Khalatnikov, I. M. and Mineev, V. P.",
    title = "{A New Type of Isotropic Cosmological Models Without Singularity}",
    doi = "10.1016/0370-2693(80)90670-X",
    journal = "Phys. Lett. B",
    volume = "91",
    pages = "99--102",
    year = "1980"
}

@article{Vilenkin:1985md,
    author = "Vilenkin, Alexander",
    title = "{Classical and Quantum Cosmology of the Starobinsky Inflationary Model}",
    reportNumber = "HUTP-85-A017",
    doi = "10.1103/PhysRevD.32.2511",
    journal = "Phys. Rev. D",
    volume = "32",
    pages = "2511",
    year = "1985"
}

@article{Comments_Starobinsky,
	doi = {10.1103/physrevd.89.043527},
	url = {https://doi.org/10.1103%2Fphysrevd.89.043527},
	year = {2014},
	month = {feb},
	publisher = {American Physical Society ({APS})},
	volume = {89},
	number = {4},
	author = {Alexandros Kehagias and Azadeh Moradinezhad Dizgah and Antonio Riotto},
	title = "{Remarks on the Starobinsky model of inflation and its descendants}",
	journal = {Physical Review D}
}

@article{review_particlephysics,
  title="{Review of particle physics}",
  author={Patrignani, Claudia and Agashe, K and Aielli, G and Amsler, C and Antonelli, M and Asner, DM and Baer, H and Banerjee, Sw and Barnett, RM and Basaglia, T and others},
  year={2016}
}

@book{Primordial,
    author = {Montani, G. and Battisti, M. V. and Benini, R. and Imponente, G.},
    title = "{Primordial cosmology}",
    publisher = "World Scientific",
    address = "Singapore",
    doi = {10.1142/7235},
    year = "2009"
}

@article{matcre_montani2001,
    author = "Montani, Giovanni",
    title = "{Influence of the particles creation on the flat and negative curved FLRW universes}",
    eprint = "gr-qc/0101113",
    archivePrefix = "arXiv",
    doi = "10.1088/0264-9381/18/1/311",
    journal = "Class. Quant. Grav.",
    volume = "18",
    pages = "193--203",
    year = "2001"
}

@article{matcre_calvaoLima,
  title={On the thermodynamics of matter creation in cosmology},
  author={Calvao, MO and Lima, JAS and Waga, I},
  journal={Physics Letters A},
  volume={162},
  number={3},
  pages={223--226},
  year={1992},
  publisher={Elsevier}
}

@article{matcre_nunes2015,
  title="{Phantom behavior via cosmological creation of particles}",
  author={Nunes, Rafael C and Pav{\'o}n, Diego},
  journal={Physical Review D},
  volume={91},
  number={6},
  pages={063526},
  year={2015},
  publisher={APS}
}

@article{matcre_nunes2016,
  title="{Gravitationally induced particle production and its impact on structure formation}",
  author={Nunes, Rafael C},
  journal={General Relativity and Gravitation},
  volume={48},
  pages={1--17},
  year={2016},
  publisher={Springer}
}

@article{Schiavone:2026agq,
    author = "Schiavone, Tiziano and De Angelis, Mariaveronica and Escamilla, Luis A. and Montani, Giovanni and Di Valentino, Eleonora",
    title = "{Revisiting the Matter Creation Process: Observational Constraints on Gravitationally Induced Dark Energy and the Hubble Tension}",
    eprint = "2601.14222",
    archivePrefix = "arXiv",
    primaryClass = "astro-ph.CO",
    month = "1",
    year = "2026"
}

@article{elizalde_odintsov,
    author = "Elizalde, Emilio and Khurshudyan, Martiros and Odintsov, Sergei D.",
    title = "{Can we learn from matter creation to solve the $H_{0}$ tension problem?}",
    eprint = "2407.20285",
    archivePrefix = "arXiv",
    primaryClass = "gr-qc",
    doi = "10.1140/epjc/s10052-024-13146-1",
    journal = "Eur. Phys. J. C",
    volume = "84",
    number = "8",
    pages = "782",
    year = "2024"
}

@ARTICLE{Bondi_1948,
       author = "Bondi, H. and Gold, T.",
        title = "{The Steady-State Theory of the Expanding Universe}",
      journal = "MNRAS",
         year = "1948",
       volume = "108",
        pages = "252",
        doi = "10.1093/mnras/108.3.252"
}

@ARTICLE{Hoyle_1948,
       author = "Hoyle, F.",
        title = "{A New Model for the Expanding Universe}",
      journal = "MNRAS",
         year = "1948",
       volume = "108",
        pages = "372",
          doi = "10.1093/mnras/108.5.372"
}

@article{Lima:1999rt,
    author = "Lima, J. A. S. and Alcaniz, J. S.",
    title = "{Flat FRW cosmologies with adiabatic matter creation: Kinematic tests}",
    eprint = "astro-ph/9902337",
    archivePrefix = "arXiv",
    journal = "Astron. Astrophys.",
    volume = "348",
    pages = "1--7",
    year = "1999"
}

@ARTICLE{Singh_2011,
       author = {{Singh}, C.~P. and {Beesham}, A.},
        title = "{Early universe cosmology with particle creation: kinematics tests}",
      journal = {Astronomy and Space Science},
         year = 2011,
        month = dec,
       volume = {336},
       number = {2},
        pages = {469-477},
          doi = {10.1007/s10509-011-0781-z},
       adsurl = {https://ui.adsabs.harvard.edu/abs/2011Ap&SS.336..469S}
}

@article{Ramos_2014,
  title = {Matter creation and cosmic acceleration},
  author = {Ramos, Rudnei O. and Vargas dos Santos, Marcelo and Waga, Ioav},
  journal = {Phys. Rev. D},
  volume = {89},
  issue = {8},
  pages = {083524},
  numpages = {10},
  year = {2014},
  month = {Apr},
  publisher = {American Physical Society},
  doi = {10.1103/PhysRevD.89.083524},
  url = {https://link.aps.org/doi/10.1103/PhysRevD.89.083524}
}

@article{deHaro:2015hdp,
    author = "de Haro, Jaume and Pan, Supriya",
    title = "{Gravitationally induced adiabatic particle production: From Big Bang to de Sitter}",
    eprint = "1512.03100",
    archivePrefix = "arXiv",
    primaryClass = "gr-qc",
    doi = "10.1088/0264-9381/33/16/165007",
    journal = "Class. Quant. Grav.",
    volume = "33",
    number = "16",
    pages = "165007",
    year = "2016"
}

@article{Zeldovich:1971mw,
    author = "Zel'dovich, Ya. B. and Starobinsky, Alexei A.",
    title = "{Particle Production and Vacuum Polarization in an Anisotropic Gravitational Field}",
    journal = "Sov. Phys. JETP",
    volume = "34",
    number = "6",
    pages = "1159--1166",
    year = "1972"
}

@article{Parker:1968mv,
    author = "Parker, L.",
    title = "{Particle creation in expanding universes}",
    doi = "10.1103/PhysRevLett.21.562",
    journal = "Phys. Rev. Lett.",
    volume = "21",
    pages = "562--564",
    year = "1968"
}

@article{Ford:1986sy,
    author = "Ford, L. H.",
    title = "{Gravitational Particle Creation and Inflation}",
    reportNumber = "TUTP-86-8",
    doi = "10.1103/PhysRevD.35.2955",
    journal = "Phys. Rev. D",
    volume = "35",
    pages = "2955",
    year = "1987"
}

@article{Kolb:2023ydq,
    author = "Kolb, Edward W. and Long, Andrew J.",
    title = "{Cosmological gravitational particle production and its implications for cosmological relics}",
    eprint = "2312.09042",
    archivePrefix = "arXiv",
    primaryClass = "astro-ph.CO",
    doi = "10.1103/RevModPhys.96.045005",
    journal = "Rev. Mod. Phys.",
    volume = "96",
    number = "4",
    pages = "045005",
    year = "2024"
}

@article{Ford:2021syk,
    author = "Ford, L. H.",
    title = "{Cosmological particle production: a review}",
    eprint = "2112.02444",
    archivePrefix = "arXiv",
    primaryClass = "gr-qc",
    doi = "10.1088/1361-6633/ac1b23",
    journal = "Rept. Prog. Phys.",
    volume = "84",
    number = "11",
    pages = "116901",
    year = "2021"
}

@article{Sotiriou_rev,
    author = "Sotiriou, Thomas P. and Faraoni, Valerio",
    title = "{f(R) Theories Of Gravity}",
    eprint = "0805.1726",
    archivePrefix = "arXiv",
    primaryClass = "gr-qc",
    doi = "10.1103/RevModPhys.82.451",
    journal = "Rev. Mod. Phys.",
    volume = "82",
    pages = "451--497",
    year = "2010"
}

@article{Olmo:2005zr,
    author = "Olmo, Gonzalo J.",
    title = "{The Gravity Lagrangian according to solar system experiments}",
    eprint = "gr-qc/0505101",
    archivePrefix = "arXiv",
    doi = "10.1103/PhysRevLett.95.261102",
    journal = "Phys. Rev. Lett.",
    volume = "95",
    pages = "261102",
    year = "2005"
}

@article{Sotiriou,
    author = "Sotiriou, Thomas P.",
    title = "{f(R) gravity and scalar-tensor theory}",
    eprint = "gr-qc/0604028",
    archivePrefix = "arXiv",
    doi = "10.1088/0264-9381/23/17/003",
    journal = "Class. Quant. Grav.",
    volume = "23",
    pages = "5117--5128",
    year = "2006"
}

@article{Beyond,
    author = "Faraoni, Valerio and Capozziello, Salvatore",
    title = "{Beyond Einstein Gravity}: {A Survey of Gravitational Theories for Cosmology and Astrophysics}",
    doi = "10.1007/978-94-007-0165-6",
    isbn = "978-94-007-0164-9, 978-94-007-0165-6",
    publisher = "Springer",
    address = "Dordrecht",
    year = "2011"
}

@article{montani_mary_fR,
    author = "Montani, Giovanni and De Angelis, Mariaveronica and Bombacigno, Flavio and Carlevaro, Nakia",
    title = "{Metric f(R) gravity with dynamical dark energy as a scenario for the Hubble tension}",
    eprint = "2306.11101",
    archivePrefix = "arXiv",
    primaryClass = "gr-qc",
    doi = "10.1093/mnrasl/slad159",
    journal = "Mon. Not. Roy. Astron. Soc.",
    volume = "527",
    number = "1",
    pages = "L156--L161",
    year = "2023"
}

@article{nojiri_2017,
  title="{Modified gravity theories on a nutshell: Inflation, bounce and late-time evolution}",
  author={Nojiri, Sh and Odintsov, SD and Oikonomou, VK3683913},
  journal={Physics Reports},
  volume={692},
  pages={1--104},
  year={2017},
  publisher={Elsevier}
}

@article{desicollaboration2025desidr2resultsii,
    author = "Abdul Karim, M. and others",
    collaboration = "DESI",
    title = "{DESI DR2 Results II: Measurements of Baryon Acoustic Oscillations and Cosmological Constraints}",
    eprint = "2503.14738",
    archivePrefix = "arXiv",
    primaryClass = "astro-ph.CO",
    reportNumber = "FERMILAB-PUB-25-0169-PPD",
    month = "3",
    year = "2025"
}

@article{BBN,
    author = {Sch{\"o}neberg, Nils},
    title = "{The 2024 BBN baryon abundance update}",
    eprint = "2401.15054",
    archivePrefix = "arXiv",
    primaryClass = "astro-ph.CO",
    doi = "10.1088/1475-7516/2024/06/006",
    journal = "JCAP",
    volume = "06",
    pages = "006",
    year = "2024"
}

@article{Cobaya,
       author = {{Torrado}, Jes{\'u}s and {Lewis}, Antony},
        title = "{Cobaya: Bayesian analysis in cosmology}",
 howpublished = {Astrophysics Source Code Library, record ascl:1910.019},
         year = 2019,
        month = oct,
          eid = {ascl:1910.019},
        pages = {ascl:1910.019},
archivePrefix = {ascl},
       eprint = {1910.019},
       adsurl = {https://ui.adsabs.harvard.edu/abs/2019ascl.soft10019T}
}

@article{Lewis:1999bs,
      author = {{Lewis}, Antony and {Challinor}, Anthony and {Lasenby}, Anthony},
        title = "{Efficient Computation of Cosmic Microwave Background Anisotropies in Closed Friedmann-Robertson-Walker Models}",
      journal = {The Astrophysical Journal},
         year = 2000,
        month = aug,
       volume = {538},
       number = {2},
        pages = {473-476},
          doi = {10.1086/309179},
archivePrefix = {arXiv},
       eprint = {astro-ph/9911177},
 primaryClass = {astro-ph},
       adsurl = {https://ui.adsabs.harvard.edu/abs/2000ApJ...538..473L}
}

@article{DESIDR1,
    author = "Adame, A. G. and others",
    collaboration = "DESI",
    title = "{DESI 2024 VI: cosmological constraints from the measurements of baryon acoustic oscillations}",
    eprint = "2404.03002",
    archivePrefix = "arXiv",
    primaryClass = "astro-ph.CO",
    reportNumber = "FERMILAB-PUB-24-0154-PPD",
    doi = "10.1088/1475-7516/2025/02/021",
    journal = "JCAP",
    volume = "02",
    pages = "021",
    year = "2025"
}

@article{Brout_2022,
    author = "Brout, Dillon and others",
    title = "{The Pantheon+ Analysis: Cosmological Constraints}",
    eprint = "2202.04077",
    archivePrefix = "arXiv",
    primaryClass = "astro-ph.CO",
    doi = "10.3847/1538-4357/ac8e04",
    journal = "Astrophys. J.",
    volume = "938",
    number = "2",
    pages = "110",
    year = "2022"
}

@article{Riess_2022,
    author = "Riess, Adam G. and others",
    title = "{A Comprehensive Measurement of the Local Value of the Hubble Constant with 1 km s$^{-1}$ Mpc$^{-1}$ Uncertainty from the Hubble Space Telescope and the SH0ES Team}",
    eprint = "2112.04510",
    archivePrefix = "arXiv",
    primaryClass = "astro-ph.CO",
    doi = "10.3847/2041-8213/ac5c5b",
    journal = "Astrophys. J. Lett.",
    volume = "934",
    number = "1",
    pages = "L7",
    year = "2022"
}

@article{bayesian,
  title = {{LCDM and Beyond: Cosmology Tools in Theory and in Practice: "Statistics and model selection in cosmology"}},
  author= "Signe Riemer-Sorensen",
  year = 2018,
  howpublished =  {\url{http://icg.port.ac.uk/~jschewts/cantata/L5/Statistics_Notes.pdf}},
  note = {Accessed: 2018-05-18}
}

@article{Zhao_2017,
    author = {Zhao, Ming-Ming and Li, Yun-He and Zhang, Jing-Fei and Zhang, Xin},
    title = "{Constraining neutrino mass and extra relativistic degrees of freedom in dynamical dark energy models using Planck 2015 data in combination with low-redshift cosmological probes: basic extensions to $\Lambda$CDM cosmology}",
    doi = "10.1093/mnras/stx978",
    journal={Monthly Notices of the Royal Astronomical Society},
   publisher={Oxford University Press (OUP)},
    volume={469},
    year = "2017"
}

@article{DeFelice:2010aj,
    author = "De Felice, Antonio and Tsujikawa, Shinji",
    title = "{f(R) theories}",
    eprint = "1002.4928",
    archivePrefix = "arXiv",
    primaryClass = "gr-qc",
    doi = "10.12942/lrr-2010-3",
    journal = "Living Rev. Rel.",
    volume = "13",
    pages = "3",
    year = "2010"
}

@article{Hwang:2001pu,
    author = "Hwang, Jai-chan and Noh, Hyerim",
    title = "{f(R) gravity theory and CMBR constraints}",
    eprint = "astro-ph/0102423",
    archivePrefix = "arXiv",
    doi = "10.1016/S0370-2693(01)00404-X",
    journal = "Phys. Lett. B",
    volume = "506",
    pages = "13--19",
    year = "2001"
}

@article{Hwang:1996xh,
    author = "Hwang, Jai-chan and Noh, Hyerim",
    title = "{Cosmological perturbations in generalized gravity theories}",
    reportNumber = "PRINT-96-116 (KYUNGPOOK)",
    doi = "10.1103/PhysRevD.54.1460",
    journal = "Phys. Rev. D",
    volume = "54",
    pages = "1460--1473",
    year = "1996"
}

@article{Ivanov:2021chn,
    author = "Ivanov, Vsevolod R. and Ketov, Sergei V. and Pozdeeva, Ekaterina O. and Vernov, Sergey Yu.",
    title = "{Analytic extensions of Starobinsky model of inflation}",
    eprint = "2111.09058",
    archivePrefix = "arXiv",
    primaryClass = "gr-qc",
    reportNumber = "IPMU21-0073",
    doi = "10.1088/1475-7516/2022/03/058",
    journal = "JCAP",
    volume = "03",
    number = "03",
    pages = "058",
    year = "2022"
}

@article{Ivanov:2025nsx,
    author = "Ivanov, Vsevolod R.",
    title = "{Inflationary Slow-Roll Parameters in the Jordan Frame for Cosmological F(R) Gravity Models}",
    eprint = "2508.14250",
    archivePrefix = "arXiv",
    primaryClass = "gr-qc",
    month = "8",
    year = "2025"
}

@article{SPT-3G:2025bzu,
    author = "Camphuis, E. and others",
    collaboration = "SPT-3G",
    title = "{SPT-3G D1: CMB temperature and polarization power spectra and cosmology from 2019 and 2020 observations of the SPT-3G Main field}",
    eprint = "2506.20707",
    archivePrefix = "arXiv",
    primaryClass = "astro-ph.CO",
    reportNumber = "FERMILAB-PUB-25-0144-PPD",
    month = "6",
    year = "2025"
}

@article{Tristram:2021tvh,
    author = "Tristram, M. and others",
    title = "{Improved limits on the tensor-to-scalar ratio using BICEP and Planck data}",
    eprint = "2112.07961",
    archivePrefix = "arXiv",
    primaryClass = "astro-ph.CO",
    doi = "10.1103/PhysRevD.105.083524",
    journal = "Phys. Rev. D",
    volume = "105",
    number = "8",
    pages = "083524",
    year = "2022"
}

@article{BICEP:2021xfz,
    author = "Ade, P. A. R. and others",
    collaboration = "BICEP, Keck",
    title = "{Improved Constraints on Primordial Gravitational Waves using Planck, WMAP, and BICEP/Keck Observations through the 2018 Observing Season}",
    eprint = "2110.00483",
    archivePrefix = "arXiv",
    primaryClass = "astro-ph.CO",
    doi = "10.1103/PhysRevLett.127.151301",
    journal = "Phys. Rev. Lett.",
    volume = "127",
    number = "15",
    pages = "151301",
    year = "2021"
}

@article{DiMarco:2024yzn,
    author = "Di Marco, Alessandro Di and Orazi, Emanuele and Pradisi, Gianfranco",
    title = "{Introduction to the Number of e-Folds in Slow-Roll Inflation}",
    eprint = "2408.01854",
    archivePrefix = "arXiv",
    primaryClass = "astro-ph.CO",
    doi = "10.3390/universe10070284",
    journal = "Universe",
    volume = "10",
    number = "7",
    pages = "284",
    year = "2024"
}

@article{Liddle:2003as,
    author = "Liddle, Andrew R and Leach, Samuel M",
    title = "{How long before the end of inflation were observable perturbations produced?}",
    eprint = "astro-ph/0305263",
    archivePrefix = "arXiv",
    doi = "10.1103/PhysRevD.68.103503",
    journal = "Phys. Rev. D",
    volume = "68",
    pages = "103503",
    year = "2003"
}

@article{German:2022sjd,
    author = "Germ{\'a}n, Gabriel and Quaglia, R. Gonzalez and Colorado, A. M. Moran",
    title = "{Model independent bounds for the number of e-folds during the evolution of the universe}",
    eprint = "2212.03730",
    archivePrefix = "arXiv",
    primaryClass = "gr-qc",
    doi = "10.1088/1475-7516/2023/03/004",
    journal = "JCAP",
    volume = "03",
    pages = "004",
    year = "2023"
}

@article{Remmen:2014mia,
    author = "Remmen, Grant N. and Carroll, Sean M.",
    title = "{How Many $e$-Folds Should We Expect from High-Scale Inflation?}",
    eprint = "1405.5538",
    archivePrefix = "arXiv",
    primaryClass = "hep-th",
    reportNumber = "CALT-2014-138",
    doi = "10.1103/PhysRevD.90.063517",
    journal = "Phys. Rev. D",
    volume = "90",
    number = "6",
    pages = "063517",
    year = "2014"
}

@article{Dolgov:2003px,
    author = "Dolgov, A. D. and Kawasaki, Masahiro",
    title = "{Can modified gravity explain accelerated cosmic expansion?}",
    eprint = "astro-ph/0307285",
    archivePrefix = "arXiv",
    doi = "10.1016/j.physletb.2003.08.039",
    journal = "Phys. Lett. B",
    volume = "573",
    pages = "1--4",
    year = "2003"
}

@article{Appleby:2009uf,
    author = "Appleby, Stephen A. and Battye, Richard A. and Starobinsky, Alexei A.",
    title = "{Curing singularities in cosmological evolDeFelice:2010ajution of F(R) gravity}",
    eprint = "0909.1737",
    archivePrefix = "arXiv",
    primaryClass = "astro-ph.CO",
    doi = "10.1088/1475-7516/2010/06/005",
    journal = "JCAP",
    volume = "06",
    pages = "005",
    year = "2010"
}

@article{Dorsch:2024nan,
    author = "Dorsch, Gl{\'a}uber C. and Miranda, Luiz and Yokomizo, Nelson",
    title = "{Gravitational reheating in Starobinsky inflation}",
    eprint = "2406.04161",
    archivePrefix = "arXiv",
    primaryClass = "gr-qc",
    doi = "10.1088/1475-7516/2024/11/050",
    journal = "JCAP",
    volume = "11",
    pages = "050",
    year = "2024"
}

@article{Motohashi:2012tt,
    author = "Motohashi, Hayato and Nishizawa, Atsushi",
    title = "{Reheating after f(R) inflation}",
    eprint = "1204.1472",
    archivePrefix = "arXiv",
    primaryClass = "astro-ph.CO",
    reportNumber = "RESCEU-8-12",
    doi = "10.1103/PhysRevD.86.083514",
    journal = "Phys. Rev. D",
    volume = "86",
    pages = "083514",
    year = "2012"
}
\end{document}